
\input epsf
\input harvmac.tex

\def\pants{1}
\def\openmult{2}
\def\openchannel{3}
\def\closedchannel{4}
\def\anomcancel{5}
\def\Dinstab{6}
\def\Dviol{7}
\def\Dviolii{8}
\def\Dvioliii{9}
\def\symmbrane{10}
\def\Ddistance{11}
\def\rotated{12}




\def\IL{\relax{\rm I\kern-.18em L}}
\def\IH{\relax{\rm I\kern-.18em H}}
\def\IR{\relax{\rm I\kern-.18em R}}
\def\IC{\relax\hbox{$\inbar\kern-.3em{\rm C}$}}
\def\IZ{\relax\ifmmode\mathchoice
{\hbox{\cmss Z\kern-.4em Z}}{\hbox{\cmss Z\kern-.4em Z}}
{\lower.9pt\hbox{\cmsss Z\kern-.4em Z}}
{\lower1.2pt\hbox{\cmsss Z\kern-.4em Z}}\else{\cmss Z\kern-.4em Z}\fi}
\def\CA{{\cal A}}
\def\CB {{\cal B}}
\def\CC {{\cal C}}

\def\CN {{\cal N}}

\def\CF {{\cal F}}
\def\CP {{\cal P }}
\def\CL {{\cal L}}

\def\CO {{\cal O}}
\def\CW {{\cal W}}

\def\CH {{\cal H}}
\def\CS {{\cal S}}

\font\manual=manfnt \def\dbend{\lower3.5pt\hbox{\manual\char127}}

\def\c{\cdot}
\def\IZ{\relax\ifmmode\mathchoice
{\hbox{\cmss Z\kern-.4em Z}}{\hbox{\cmss Z\kern-.4em Z}}
{\lower.9pt\hbox{\cmsss Z\kern-.4em Z}}
{\lower1.2pt\hbox{\cmsss Z\kern-.4em Z}}\else{\cmss Z\kern-.4em Z}\fi}
\def\half {{1\over 2}}

\def\p{\partial}
\def\pb{\bar{\partial}}

\def\CL {{\cal L}}

\def\CN {{\cal N}}

\def\CO {{\cal O}}
\def\CJ {{\cal J}}

\def\CP {{\cal P }}

\def\CS {{\cal S }}


\def\IZ{\relax\ifmmode\mathchoice
{\hbox{\cmss Z\kern-.4em Z}}{\hbox{\cmss Z\kern-.4em Z}}
{\lower.9pt\hbox{\cmsss Z\kern-.4em Z}}
{\lower1.2pt\hbox{\cmsss Z\kern-.4em Z}}\else{\cmss Z\kern-.4em
Z}\fi}
\def\IB{\relax{\rm I\kern-.18em B}}
\def\IC{{\relax\hbox{$\inbar\kern-.3em{\rm C}$}}}
\def\ID{\relax{\rm I\kern-.18em D}}
\def\IE{\relax{\rm I\kern-.18em E}}
\def\IF{\relax{\rm I\kern-.18em F}}
\def\IG{\relax\hbox{$\inbar\kern-.3em{\rm G}$}}
\def\IGa{\relax\hbox{${\rm I}\kern-.18em\Gamma$}}
\def\IH{\relax{\rm I\kern-.18em H}}
\def\II{\relax{\rm I\kern-.18em I}}
\def\IK{\relax{\rm I\kern-.18em K}}
\def\IP{\relax{\rm I\kern-.18em P}}

\def\IQ{\relax\hbox{$\inbar\kern-.3em{\rm Q}$}}
\def\IP{\relax{\rm I\kern-.18em P}}

\def\IB{\relax{\rm I\kern-.18em B}}
\def\IC{\Bbb{C} }
\def\ID{\relax{\rm I\kern-.18em D}}
\def\IE{\relax{\rm I\kern-.18em E}}
\def\IF{\relax{\rm I\kern-.18em F}}
\def\IG{\relax\hbox{$\inbar\kern-.3em{\rm G}$}}
\def\IGa{\relax\hbox{${\rm I}\kern-.18em\Gamma$}}
\def\IH{\relax{\rm I\kern-.18em H}}
\def\II{\relax{\rm I\kern-.18em I}}
\def\IJ{\relax{\rm I\kern-.18em J}}
\def\IK{\relax{\rm I\kern-.18em K}}
\def\IL{\relax{\rm I\kern-.18em L}}

\def\IN{\relax{\rm I\kern-.18em N}}
\def\IO{\relax{\rm I\kern-.18em O}}
\def\IP{\relax{\rm I\kern-.18em P}}
\def\IQ{\relax\hbox{$\inbar\kern-.3em{\rm Q}$}}
\def\IR{\relax{\rm I\kern-.18em R}}
\def\IW{\relax\hbox{$\inbar\kern-.3em{\rm W}$}}

\def\liet{{\underline{\bf t}}}

\def\inbar{\,\vrule height1.5ex width.4pt depth0pt}
\def\mod{\rm mod}
\def\ndt{\noindent}
\def\p{\partial} 
\def\pb{{\bar \p}}

\font\cmss=cmss10 \font\cmsss=cmss10 at 7pt
\def\IR{\relax{\rm I\kern-.18em R}}

\def\Tr{\rm Tr} 
\def\vol{{\rm vol}}

\def\bgn{\bigskip\ndt}
\def\IC{{\bf C}}


%
\def\inv{^{\raise.15ex\hbox{${\scriptscriptstyle -}$}\kern-.05em 1}}

\def\Dsl{\,\raise.15ex\hbox{/}\mkern-13.5mu D} 
\def\dsl{\raise.15ex\hbox{/}\kern-.57em\partial}

 \def\Tr{{\rm Tr}}


\def\rara{\rangle\rangle}



\def\boxit#1{\vbox{\hrule\hbox{\vrule\kern8pt
\vbox{\hbox{\kern8pt}\hbox{\vbox{#1}}\hbox{\kern8pt}}
\kern8pt\vrule}\hrule}}
\def\mathboxit#1{\vbox{\hrule\hbox{\vrule\kern8pt\vbox{\kern8pt
\hbox{$\displaystyle #1$}\kern8pt}\kern8pt\vrule}\hrule}}

\lref\atiyah{M.F. Atiyah, ``Global theory of 
elliptic operators,''  Proc. Int. Conf. Func. 
Anal. and Related Topics, pp.  21-30, 
Univ. of Tokyo Press, Tokyo, 1970}

\lref\segalitp{See lectures by G. Segal 
at http:// online.kitp.ucsb.edu/ online/geom99, and lecture notes 
at  http:// www.cgtp.duke.edu/ITP99/segal/}

\lref\donaldson{S.K. Donaldson and P.B. Kronheimer, 
{\it The Geometry of Four-Manifolds}, Oxford, 1990 }

\lref\BouwknegtVU{
P.~Bouwknegt, A.~L.~Carey, V.~Mathai, M.~K.~Murray and D.~Stevenson,
``Twisted K-theory and K-theory of bundle gerbes,''
Commun.\ Math.\ Phys.\  {\bf 228}, 17 (2002)
[arXiv:hep-th/0106194].
}

\lref\BouwknegtQT{
P.~Bouwknegt and V.~Mathai,
``D-branes, B-fields and twisted K-theory,''
JHEP {\bf 0003}, 007 (2000)
[arXiv:hep-th/0002023].
}

\lref\fqs{D. Friedan, Z. Qiu, and S.H. Shenker,
Phys. Lett. {\bf 151B}(1985)37}
\lref\knizhnik{P. di Vecchia, V.G. Knizhnik, J.L. Petersen, and
P. Rossi, Nucl. Phys. {\bf B253}(1985) 701}
\lref\kac{V.G. Kac and I.T. Todorov, Comm. Math. Phys.
{\bf 102}(1985)337}
%

\lref\AsakawaUI{
T.~Asakawa, S.~Sugimoto and S.~Terashima,
``Exact description of D-branes via tachyon condensation,''
arXiv:hep-th/0212188.
}
\lref\AsakawaVM{
T.~Asakawa, S.~Sugimoto and S.~Terashima,
``D-branes, matrix theory and K-homology,''
JHEP {\bf 0203}, 034 (2002)
[arXiv:hep-th/0108085].
}

\lref\AsakawaUI{
T.~Asakawa, S.~Sugimoto and S.~Terashima,
``Exact description of D-branes via tachyon condensation,''
arXiv:hep-th/0212188.
}

\lref\BordaloEC{
P.~Bordalo, S.~Ribault and C.~Schweigert,
``Flux stabilization in compact groups,''
JHEP {\bf 0110}, 036 (2001)
[arXiv:hep-th/0108201].
}

\lref\BachasIK{
C.~Bachas, M.~R.~Douglas and C.~Schweigert,
``Flux stabilization of D-branes,''
JHEP {\bf 0005}, 048 (2000)
[arXiv:hep-th/0003037].
}

\lref\SchweigertIX{
C.~Schweigert, J.~Fuchs and J.~Walcher,
``Conformal field theory, boundary conditions and applications to string  theory,''
arXiv:hep-th/0011109.
}

\lref\PawelczykAH{
J.~Pawelczyk,
``SU(2) WZW D-branes and their noncommutative geometry from DBI action,''
JHEP {\bf 0008}, 006 (2000)
[arXiv:hep-th/0003057].
}

\lref\FelderKA{
G.~Felder, J.~Frohlich, J.~Fuchs and C.~Schweigert,
``The geometry of WZW branes,''
J.\ Geom.\ Phys.\  {\bf 34}, 162 (2000)
[arXiv:hep-th/9909030].
}

\lref\MartinecWG{
E.~J.~Martinec and G.~Moore,
``On decay of K-theory,''
arXiv:hep-th/0212059.
}

\lref\MaldacenaSS{
J.~M.~Maldacena, G.~W.~Moore and N.~Seiberg,
``D-brane charges in five-brane backgrounds,''
JHEP {\bf 0110}, 005 (2001)
[arXiv:hep-th/0108152].
}

\lref\MaldacenaXJ{
J.~M.~Maldacena, G.~W.~Moore and N.~Seiberg,
``D-brane instantons and K-theory charges,''
JHEP {\bf 0111}, 062 (2001)
[arXiv:hep-th/0108100].
}

\lref\MaldacenaKY{
J.~M.~Maldacena, G.~W.~Moore and N.~Seiberg,
``Geometrical interpretation of D-branes in gauged WZW models,''
JHEP {\bf 0107}, 046 (2001)
[arXiv:hep-th/0105038].
}

\lref\KutasovAQ{
D.~Kutasov, M.~Marino and G.~W.~Moore,
``Remarks on tachyon condensation in superstring field theory,''
arXiv:hep-th/0010108.
}

\lref\KutasovQP{
D.~Kutasov, M.~Marino and G.~W.~Moore,
``Some exact results on tachyon condensation in string field theory,''
JHEP {\bf 0010}, 045 (2000)
[arXiv:hep-th/0009148].
}

\lref\HarveyTE{
J.~A.~Harvey and G.~W.~Moore,
``Noncommutative tachyons and K-theory,''
J.\ Math.\ Phys.\  {\bf 42}, 2765 (2001)
[arXiv:hep-th/0009030].
}

\lref\MinasianMM{
R.~Minasian and G.~W.~Moore,
``K-theory and Ramond-Ramond charge,''
JHEP {\bf 9711}, 002 (1997)
[arXiv:hep-th/9710230].
}

\lref\GreenDD{
M.~B.~Green, J.~A.~Harvey and G.~W.~Moore,
``I-brane inflow and anomalous couplings on D-branes,''
Class.\ Quant.\ Grav.\  {\bf 14}, 47 (1997)
[arXiv:hep-th/9605033].
}

\lref\MartinecTZ{
E.~J.~Martinec,
``Defects, decay, and dissipated states,''
arXiv:hep-th/0210231.
}

\lref\HarveyNA{
J.~A.~Harvey, D.~Kutasov and E.~J.~Martinec,
``On the relevance of tachyons,''
arXiv:hep-th/0003101.
}

\lref\KutasovAQ{
D.~Kutasov, M.~Marino and G.~W.~Moore,
``Remarks on tachyon condensation in superstring field theory,''
arXiv:hep-th/0010108.
}

\lref\SzaboJV{
R.~J.~Szabo,
``D-branes, tachyons and K-homology,''
Mod.\ Phys.\ Lett.\ A {\bf 17}, 2297 (2002)
[arXiv:hep-th/0209210].
}

\lref\SzaboYD{
R.~J.~Szabo,
``Superconnections, anomalies and non-BPS brane charges,''
J.\ Geom.\ Phys.\  {\bf 43}, 241 (2002)
[arXiv:hep-th/0108043].
}

\lref\OlsenXX{
K.~Olsen and R.~J.~Szabo,
``Constructing D-branes from K-theory,''
Adv.\ Theor.\ Math.\ Phys.\  {\bf 3}, 889 (1999)
[arXiv:hep-th/9907140].
}

\lref\WittenCN{
E.~Witten,
``Overview of K-theory applied to strings,''
Int.\ J.\ Mod.\ Phys.\ A {\bf 16}, 693 (2001)
[arXiv:hep-th/0007175].
}

\lref\WittenNZ{
E.~Witten,
``Noncommutative tachyons and string field theory,''
arXiv:hep-th/0006071.
}

\lref\DiaconescuWZ{
D.~E.~Diaconescu, G.~W.~Moore and E.~Witten,
``A derivation of K-theory from M-theory,''
arXiv:hep-th/0005091.
}

\lref\DiaconescuWY{
D.~E.~Diaconescu, G.~W.~Moore and E.~Witten,
``E(8) gauge theory, and a derivation of K-theory from M-theory,''
arXiv:hep-th/0005090.
}

\lref\FreedVC{
D.~S.~Freed and E.~Witten,
``Anomalies in string theory with D-branes,''
arXiv:hep-th/9907189.
}

\lref\WittenCD{
E.~Witten,
``D-branes and K-theory,''
JHEP {\bf 9812}, 019 (1998)
[arXiv:hep-th/9810188].
}

\lref\KapustinDI{
A.~Kapustin,
``D-branes in a topologically nontrivial B-field,''
Adv.\ Theor.\ Math.\ Phys.\  {\bf 4}, 127 (2000)
[arXiv:hep-th/9909089].
}

\lref\HikidaPY{
Y.~Hikida, M.~Nozaki and Y.~Sugawara,
``Formation of spherical D2-brane from multiple D0-branes,''
Nucl.\ Phys.\ B {\bf 617}, 117 (2001)
[arXiv:hep-th/0101211].
}

\lref\SchomerusDC{
V.~Schomerus,
``Lectures on branes in curved backgrounds,''
Class.\ Quant.\ Grav.\  {\bf 19}, 5781 (2002)
[arXiv:hep-th/0209241].
}

\lref\AlekseevWG{
A.~Y.~Alekseev, A.~Recknagel and V.~Schomerus,
``Open strings and non-commutative geometry of branes on group manifolds,''
Mod.\ Phys.\ Lett.\ A {\bf 16}, 325 (2001)
[arXiv:hep-th/0104054].
}

\lref\FredenhagenEI{
S.~Fredenhagen and V.~Schomerus,
``Branes on group manifolds, gluon condensates, and twisted K-theory,''
JHEP {\bf 0104}, 007 (2001)
[arXiv:hep-th/0012164].
}

\lref\AlekseevJX{
A.~Alekseev and V.~Schomerus,
``RR charges of D2-branes in the WZW model,''
arXiv:hep-th/0007096.
}

\lref\AlekseevFD{
A.~Y.~Alekseev, A.~Recknagel and V.~Schomerus,
``Brane dynamics in background fluxes and non-commutative geometry,''
JHEP {\bf 0005}, 010 (2000)
[arXiv:hep-th/0003187].
}

\lref\mooresegal{G. Moore and G. Segal, unpublished. 
The material is available at http://  
online.kitp.ucsb.edu/
online/mp01/moore1/
and  /moore2, and from lecture notes at the Clay Institute School 
on Geometry and String Theory, held at the 
Isaac Newton Institute, Cambridge, UK. April 2002. } 

\lref\AffleckTK{
I.~Affleck and A.~W.~Ludwig,
``Universal Noninteger 'Ground State Degeneracy' In Critical Quantum Systems,''
Phys.\ Rev.\ Lett.\  {\bf 67}, 161 (1991).
}
\lref\AffleckNG{
I.~Affleck and A.~W.~Ludwig,
``Exact Conformal Field Theory Results On The Multichannel Kondo Effect: Single Fermion Green's Function, Selfenergy And Resistivity,''
UBCTP-92-029
}

\lref\KrausNJ{
P.~Kraus and F.~Larsen,
``Boundary string field theory of the DD-bar system,''
Phys.\ Rev.\ D {\bf 63}, 106004 (2001)
[arXiv:hep-th/0012198].
}

\lref\TakayanagiRZ{
T.~Takayanagi, S.~Terashima and T.~Uesugi,
``Brane-antibrane action from boundary string field theory,''
JHEP {\bf 0103}, 019 (2001)
[arXiv:hep-th/0012210].
}

\lref\WittenQY{
E.~Witten,
``On background independent open string field theory,''
Phys.\ Rev.\ D {\bf 46}, 5467 (1992)
[arXiv:hep-th/9208027].
}

\lref\WittenCR{
E.~Witten,
``Some computations in background independent off-shell string theory,''
Phys.\ Rev.\ D {\bf 47}, 3405 (1993)
[arXiv:hep-th/9210065].
}

\lref\ShatashviliUX{
S.~L.~Shatashvili,
``On field theory of open strings, tachyon condensation and closed  strings,''
arXiv:hep-th/0105076.
}

\lref\GerasimovGA{
A.~A.~Gerasimov and S.~L.~Shatashvili,
``Stringy Higgs mechanism and the fate of open strings,''
JHEP {\bf 0101}, 019 (2001)
[arXiv:hep-th/0011009].
}

\lref\GerasimovZP{
A.~A.~Gerasimov and S.~L.~Shatashvili,
``On exact tachyon potential in open string field theory,''
JHEP {\bf 0010}, 034 (2000)
[arXiv:hep-th/0009103].
}

\lref\ShatashviliPS{
S.~L.~Shatashvili,
``On the problems with background independence in string theory,''
arXiv:hep-th/9311177.
}

\lref\ShatashviliKK{
S.~L.~Shatashvili,
``Comment on the background independent open string theory,''
Phys.\ Lett.\ B {\bf 311}, 83 (1993)
[arXiv:hep-th/9303143].
}

\lref\TseytlinDJ{
A.~A.~Tseytlin,
``Born-Infeld action, supersymmetry and string theory,''
arXiv:hep-th/9908105.
}

\lref\AffleckGE{
I.~Affleck,
``Conformal Field Theory Approach to the Kondo Effect,''
Acta Phys.\ Polon.\ B {\bf 26}, 1869 (1995)
[arXiv:cond-mat/9512099].
}

\lref\HarveyGQ{
J.~A.~Harvey, S.~Kachru, G.~W.~Moore and E.~Silverstein,
``Tension is dimension,''
JHEP {\bf 0003}, 001 (2000)
[arXiv:hep-th/9909072].
}

\lref\DouglasZW{
M.~R.~Douglas,
``D-branes in curved space,''
Adv.\ Theor.\ Math.\ Phys.\  {\bf 1}, 198 (1998)
[arXiv:hep-th/9703056].
}
\lref\GawedzkiSE{
K.~Gawedzki and N.~Reis,
``WZW branes and gerbes,''
arXiv:hep-th/0205233.
}


\lref\KlebanovNI{
I.~R.~Klebanov and L.~Thorlacius,
``The Size of p-Branes,''
Phys.\ Lett.\ B {\bf 371}, 51 (1996)
[arXiv:hep-th/9510200].
}

\lref\FreedMZ{
D.~S.~Freed,
``Twisted K-theory and loop groups,''
arXiv:math.at/0206237.
}

\lref\fht{D.S. Freed, M.J. Hopkins, and C. Teleman, 
``Twisted equivariant K-theory with complex coefficients,'' 
math.AT/0206257}
%

\lref\FreedJD{
D.~S.~Freed,
``The Verlinde algebra is twisted equivariant K-theory,''
arXiv:math.rt/0101038.
}

\lref\FreedTA{
D.~S.~Freed,
``Dirac charge quantization and generalized differential cohomology,''
arXiv:hep-th/0011220.
}

\lref\WittenXY{
E.~Witten,
``Baryons and branes in anti de Sitter space,''
JHEP {\bf 9807}, 006 (1998)
[arXiv:hep-th/9805112].
}
\lref\HarveyGC{
J.~A.~Harvey and G.~W.~Moore,
``On the algebras of BPS states,''
Commun.\ Math.\ Phys.\  {\bf 197}, 489 (1998)
[arXiv:hep-th/9609017].
}

\lref\LiZA{
K.~Li and E.~Witten,
``Role of short distance behavior in off-shell open string field theory,''
Phys.\ Rev.\ D {\bf 48}, 853 (1993)
[arXiv:hep-th/9303067].
}

\lref\HarveyTE{
J.~A.~Harvey and G.~W.~Moore,
``Noncommutative tachyons and K-theory,''
J.\ Math.\ Phys.\  {\bf 42}, 2765 (2001)
[arXiv:hep-th/0009030].
}

\lref\SeibergVS{
N.~Seiberg and E.~Witten,
``String theory and noncommutative geometry,''
JHEP {\bf 9909}, 032 (1999)
[arXiv:hep-th/9908142].
}

\lref\MyersPS{
R.~C.~Myers,
``Dielectric-branes,''
JHEP {\bf 9912}, 022 (1999)
[arXiv:hep-th/9910053].
}

\lref\MooreAR{
G.~W.~Moore,
``String duality, automorphic forms, and generalized Kac-Moody algebras,''
Nucl.\ Phys.\ Proc.\ Suppl.\  {\bf 67}, 56 (1998)
[arXiv:hep-th/9710198].
}

\lref\TuraevYF{
V.~Turaev,
``Homotopy field theory in dimension 2 and group-algebras,''
arXiv:  math.qa/  9910010.
}

%
\lref\HoriCK{
K.~Hori, A.~Iqbal and C.~Vafa,
``D-branes and mirror symmetry,''
arXiv:hep-th/0005247.
}
%
\lref\HoriKT{
K.~Hori and C.~Vafa,
``Mirror symmetry,''
arXiv:hep-th/0002222.
}
%
\lref\HoriFJ{
K.~Hori,
``Mirror symmetry and some applications,''
arXiv:hep-th/0106043.
}

\lref\HoriIC{
K.~Hori,
``Linear models of supersymmetric D-branes,''
arXiv:hep-th/0012179.
}
%

\lref\PeriwalEB{
V.~Periwal,
``D-brane charges and K-homology.  (Z)),''
JHEP {\bf 0007}, 041 (2000)
[arXiv:hep-th/0006223].
}

\lref\MarinoQC{
M.~Marino,
``On the BV formulation of boundary superstring field theory,''
JHEP {\bf 0106}, 059 (2001)
[arXiv:hep-th/0103089].
}

\lref\NiarchosSI{
V.~Niarchos and N.~Prezas,
``Boundary superstring field theory,''
Nucl.\ Phys.\ B {\bf 619}, 51 (2001)
[arXiv:hep-th/0103102].
}

\lref\WittenCR{
E.~Witten,
``Some computations in background independent off-shell string theory,''
Phys.\ Rev.\ D {\bf 47}, 3405 (1993)
[arXiv:hep-th/9210065].
}
\lref\GawedzkiYE{
K.~Gawedzki,
``Boundary WZW, G/H, G/G and CS theories,''
Annales Henri Poincare {\bf 3}, 847 (2002)
[arXiv:hep-th/0108044].
}
\lref\LazaroiuRK{
C.~I.~Lazaroiu,
``On the structure of open-closed topological field theory in two  dimensions,''
Nucl.\ Phys.\ B {\bf 603}, 497 (2001)
[arXiv:hep-th/0010269].
}

\lref\Connes{A. Connes, {\it Noncommutative Geometry}, 
Academic Press, 1994}. 

\lref\pressleysegal{
A. Pressley and G. Segal, {\it Loop Groups}, Oxford, 1986}

\lref\rdouglas{P. Baum and R.G. Douglas, ``K Homology and Index Theory,'' 
Proc. Symp. Pure Math. {\bf 38}(1982), 117}

\lref\WittenIM{
E.~Witten,
``Supersymmetry And Morse Theory,''
J.\ Diff.\ Geom.\  {\bf 17}, 661 (1982).
}
\lref\WittenMJ{
E.~Witten,
``Global Anomalies In String Theory,''
Print-85-0620 (PRINCETON)
{\it To appear in Proc. of Argonne Symp. on Geometry, Anomalies and Topology, Argonne, IL, Mar 28-30, 1985}
}

\lref\segalbourbaki{G. Segal, ``Elliptic cohomology,'' Asterisque {\bf 161-162}(1988)
exp. no. 695, 187-201}

\lref\stolz{S. Stolz, Contribution to this volume} 

\lref\AlbertssonDV{
C.~Albertsson, U.~Lindstrom and M.~Zabzine,
``N = 1 supersymmetric sigma model with boundaries. I,''
arXiv:hep-th/0111161.
}
\lref\AlbertssonQC{
C.~Albertsson, U.~Lindstrom and M.~Zabzine,
``N = 1 supersymmetric sigma model with boundaries. II,''
arXiv:hep-th/0202069.
}

\lref\JaffeQZ{
A.~Jaffe, A.~Lesniewski and J.~Weitsman,
``Index Of A Family Of Dirac Operators On Loop Space,''
Commun.\ Math.\ Phys.\  {\bf 112}, 75 (1987).
}

\lref\JaffePZ{
A.~Jaffe, A.~Lesniewski and K.~Osterwalder,
``Quantum K Theory. 1. The Chern Character,''
Commun.\ Math.\ Phys.\  {\bf 118}, 1 (1988).
}

\lref\ErnstBH{
K.~Ernst, P.~Feng, A.~Jaffe and A.~Lesniewski,
``Quantum K Theory. 2. Homotopy Invariance Of The Chern Character,''
HUTMP-88/B-228
}

\lref\MickelssonJX{
J.~Mickelsson,
``Gerbes, (twisted) K-theory, and the supersymmetric WZW model,''
arXiv:hep-th/0206139.
}

\lref\BazhanovFT{
V.~V.~Bazhanov, S.~L.~Lukyanov and A.~B.~Zamolodchikov,
``Integrable structure of conformal field theory, quantum KdV theory and thermodynamic Bethe ansatz,''
Commun.\ Math.\ Phys.\  {\bf 177}, 381 (1996)
[arXiv:hep-th/9412229].
}

\lref\BazhanovDR{
V.~V.~Bazhanov, S.~L.~Lukyanov and A.~B.~Zamolodchikov,
``Integrable Structure of Conformal Field Theory II. Q-operator and DDV equation,''
Commun.\ Math.\ Phys.\  {\bf 190}, 247 (1997)
[arXiv:hep-th/9604044].
}

\lref\BazhanovDQ{
V.~V.~Bazhanov, S.~L.~Lukyanov and A.~B.~Zamolodchikov,
``Integrable structure of conformal field theory. III: The Yang-Baxter  relation,''
Commun.\ Math.\ Phys.\  {\bf 200}, 297 (1999)
[arXiv:hep-th/9805008].
}

\lref\andrei{N. Andrei, ``Integrable Models in Condensed Matter Physics,''
cond-mat/9408101 }

\lref\BouwknegtBQ{
P.~Bouwknegt, P.~Dawson and D.~Ridout,
``D-branes on group manifolds and fusion rings,''
JHEP {\bf 0212}, 065 (2002)
[arXiv:hep-th/0210302].
}

\lref\StanciuVW{
S.~Stanciu,
``An illustrated guide to D-branes in SU(3),''
arXiv:hep-th/0111221.
}

\lref\FendleyKJ{
P.~Fendley, F.~Lesage and H.~Saleur,
``A unified framework for the Kondo problem and for an impurity in a Luttinger liquid,''
arXiv:cond-mat/9510055.
}

\lref\LesageQF{
F.~Lesage, H.~Saleur and P.~Simonetti,
``Boundary flows in minimal models,''
Phys.\ Lett.\ B {\bf 427}, 85 (1998)
[arXiv:hep-th/9802061].
}

\lref\SenSM{
A.~Sen,
``Tachyon condensation on the brane antibrane system,''
JHEP {\bf 9808}, 012 (1998)
[arXiv:hep-th/9805170].
}

\lref\MatsuoPJ{
Y.~Matsuo,
``Topological charges of noncommutative soliton,''
Phys.\ Lett.\ B {\bf 499}, 223 (2001)
[arXiv:hep-th/0009002].
}

\lref\MathaiIW{
V.~Mathai and I.~M.~Singer,
``Twisted K-homology theory, twisted Ext-theory,''
arXiv:hep-th/0012046.
}

\lref\FreedQP{
D.~S.~Freed,
``K-theory in quantum field theory,''
arXiv:math-ph/0206031.
}

\lref\LewellenTB{
D.~C.~Lewellen,
``Sewing constraints for conformal field theories on surfaces with boundaries,''
Nucl.\ Phys.\ B {\bf 372}, 654 (1992).
}
\lref\CardyTV{
J.~L.~Cardy and D.~C.~Lewellen,
``Bulk And Boundary Operators In Conformal Field Theory,''
Phys.\ Lett.\ B {\bf 259}, 274 (1991).
}

\lref\belovlovelace{
D. Belov and C. Lovelace, to appear} 
%
\lref\BarsAG{
I.~Bars,
``Map of Witten's * to Moyal's *,''
Phys.\ Lett.\ B {\bf 517}, 436 (2001)
[arXiv:hep-th/0106157].
}

\lref\DouglasJM{
M.~R.~Douglas, H.~Liu, G.~Moore and B.~Zwiebach,
``Open string star as a continuous Moyal product,''
JHEP {\bf 0204}, 022 (2002)
[arXiv:hep-th/0202087].
}

\lref\BarsNU{
I.~Bars and Y.~Matsuo,
``Computing in string field theory using the Moyal star product,''
Phys.\ Rev.\ D {\bf 66}, 066003 (2002)
[arXiv:hep-th/0204260].
}
\lref\BarsYJ{
I.~Bars,
``MSFT: Moyal star formulation of string field theory,''
arXiv:hep-th/0211238.
}

\lref\friedanup{D. Friedan, ``The Space of Conformal Boundary 
Conditions for the $c=1$ Gaussian Model,'' 1993, unpublished.} 
\lref\GaberdielZQ{
M.~R.~Gaberdiel and A.~Recknagel,
``Conformal boundary states for free bosons and fermions,''
JHEP {\bf 0111}, 016 (2001)
[arXiv:hep-th/0108238].
}
\lref\JanikHB{
R.~A.~Janik,
``Exceptional boundary states at c = 1,''
Nucl.\ Phys.\ B {\bf 618}, 675 (2001)
[arXiv:hep-th/0109021].
}

\lref\ConnesFE{
A.~Connes and D.~Kreimer,
``Renormalization in quantum field theory and the Riemann-Hilbert  problem. II: The beta-function, diffeomorphisms and the renormalization
group,''
Commun.\ Math.\ Phys.\  {\bf 216}, 215 (2001)
[arXiv:hep-th/0003188].
}
\lref\ConnesYR{
A.~Connes and D.~Kreimer,
``Renormalization in quantum field theory and the Riemann-Hilbert  problem. I: The Hopf algebra structure of graphs and the main theorem,''
Commun.\ Math.\ Phys.\  {\bf 210}, 249 (2000)
[arXiv:hep-th/9912092].
}

\lref\LindstromMC{
U.~Lindstrom, M.~Rocek and P.~van Nieuwenhuizen,
``Consistent boundary conditions for open strings,''
arXiv:hep-th/0211266.
}

\lref\EvslinCJ{
J.~Evslin and U.~Varadarajan,
``K-theory and S-duality: Starting over from square 3,''
arXiv:hep-th/0112084.
}
\lref\EvslinHD{
J.~Evslin,
``Twisted K-theory from monodromies,''
arXiv:hep-th/0302081.
}

\lref\FalcetoEH{
F.~Falceto and K.~Gawedzki,
``Boundary G/G theory and topological Poisson-Lie sigma model,''
Lett.\ Math.\ Phys.\  {\bf 59}, 61 (2002)
[arXiv:hep-th/0108206].
}

\lref\schafernameki{S. Schafer-Nameki, to appear.}

\lref\FredenhagenXF{
S.~Fredenhagen,
``Organizing boundary RG flows,''
arXiv:hep-th/0301229.
}
\lref\FredenhagenQN{
S.~Fredenhagen and V.~Schomerus,
``On boundary RG-flows in coset conformal field theories,''
arXiv:hep-th/0205011.
}
\lref\FredenhagenKW{
S.~Fredenhagen and V.~Schomerus,
``D-branes in coset models,''
JHEP {\bf 0202}, 005 (2002)
[arXiv:hep-th/0111189].
}
\lref\LercheIV{
W.~Lerche and J.~Walcher,
``Boundary rings and N = 2 coset models,''
Nucl.\ Phys.\ B {\bf 625}, 97 (2002)
[arXiv:hep-th/0011107].
}

\lref\DouglasGE{
M.~R.~Douglas,
``Two lectures on D-geometry and noncommutative geometry,''
arXiv:hep-th/9901146.
}

\lref\DouglasBE{
M.~R.~Douglas,
``D-branes on Calabi-Yau manifolds,''
arXiv:math.ag/0009209.
}

\lref\FuchsQC{
J.~Fuchs and C.~Schweigert,
``Category theory for conformal boundary conditions,''
arXiv:math.ct/0106050.
}

\lref\FuchsYK{
J.~Fuchs, I.~Runkel and C.~Schweigert,
``Boundaries, defects and Frobenius algebras,''
arXiv:hep-th/0302200.
}

\lref\freedtalk{D. Freed, M. Hopkins, and C. Teleman, 
presented by  D. Freed, at  ``Elliptic Cohomology and 
Chromatic Phenomena,''   Dec. 2002, Isaac Newton Institute. }

\lref\snaith{V.P. Snaith, ``On the Kunneth formula spectral 
sequence in equivariant K-theory,'' Proc. Camb.  Phil. Soc. 
{\bf 72}(1972)167} 

\lref\hodgkin{L. Hodgkin, ``The equivariant Kunneth theorem in 
K-theory,'' Springer Lect. Notes in Math, {\bf 496}, pp. 1-101, 
1975} 

\rightline{RUNHETC-2003-02 }  
\Title{  
\rightline{hep-th/0304018}}  
{\vbox{\centerline{$K$-Theory from a Physical Perspective$^*$}}}  
\bigskip  
\centerline{  Gregory Moore\footnote{}{$^*$ Summary of a 
talk delivered at the conference {\it Topology, Geometry, 
and Quantum Field Theory} Oxford, July 26, 2002}}  
 
\bigskip  
\centerline{ { Department of Physics, Rutgers University}}  
\centerline{ Piscataway, NJ 08855-0849, USA}

\bigskip  
\noindent  
This is an expository paper which aims at explaining 
a physical  point of view on the $K$-theoretic 
classification of $D$-branes. 
 We combine ideas of renormalization group 
flows between boundary conformal field theories, together 
with spacetime notions such as anomaly cancellation 
and $D$-brane instanton effects. We illustrate this point of 
view by  describing the twisted K-theory of the special 
unitary groups $SU(N)$.

\vfill  
  
\Date{March 26, 2003 }  

\newsec{Introduction}

This is an expository paper devoted to explaining 
some aspects of the $K$-theoretic classification 
of D-branes. Our aim is to address the topic in 
ways complementary to the discussions of 
  \MinasianMM\WittenCD. Reviews of the latter 
approaches include   \OlsenXX\WittenCN\FreedQP\SzaboJV. 
Our intended audience is the mathematician who is
well-versed in conformal field theory and $K$-theory, 
and has some interest in the wider universe of 
(nonconformal) quantum field theories. 

Our plan for the paper is to  
 begin in section 2 by reviewing  the relation 
of D-branes and K-theory  at the level of  topological field theory. 
Then in section 3 we will move on to discuss D-branes in conformal field theory. 
We will advocate a point of view emphasizing   
2-dimensional conformal field theories as elements  of a larger   space of 
2-dimensional quantum field theories. ``D-branes'' are identified with 
conformal quantum field theories on 2-dimensional manifolds with boundary. 
 From this vantage,  the 
topological classification of D-branes is the classification of the
connected components of the space of 
2-dimensional theories on manifolds with boundary which only break 
conformal invariance through their boundary conditions. 
  
In section 4 we will turn to conformal field theories which are used to 
build string theories. In this case, there is a spacetime 
viewpoint on the classification of D-branes. We will present a 
viewpoint on D-brane classification, based on anomaly cancellation
and ``instanton effects,'' that turns out to be closely related to the 
Atiyah-Hirzebruch spectral sequence. 

In section 5 we    examine   a detailed example, that of branes in WZW models, 
and show how, using the approach  explained in   
  sections 3 and 4  we can gain  an intuitive 
understanding of the twisted K-theory of $SU(N)$. The 
picture is in beautiful harmony with a rigorous computation 
of M. Hopkins. 

Let us warn the reader at the outset that in this modest review
 we are only  attempting 
to give a broad brush overview of some ideas. We are 
not attempting to give  a detailed and rigorous 
mathematical theory, nor are we attempting to give a comprehensive 
review of the subject.

\newsec{Branes in 2-dimensional topological field theory}

The relation of D-branes and K-theory can be 
illustrated very clearly   in the 
 extremely simple case of 2-dimensional (2D) topological field theory. 
This discussion was developed in collaboration 
with Graeme Segal  \mooresegal.

We will regard a ``field theory'' along the lines of 
Segal's contribution to this volume. It a
functor from a geometric category to some linear category.
In the simple case of 2D topological field theory the 
geometric category has as objects disjoint collections of circles
and as morphisms diffeomorphism classes of
 oriented  cobordisms between the objects.  
The target category is the category of vector spaces and linear 
transformations. 
Recall that to give a 2D topological field theory of closed strings
is to give a commutative, 
finite dimensional Frobenius algebra $\CC$. For example, the algebra 
structure follows from Fig. 1.

\bigskip
{\vbox{{\epsfxsize=1.1in
        \nobreak
    \centerline{\epsfbox{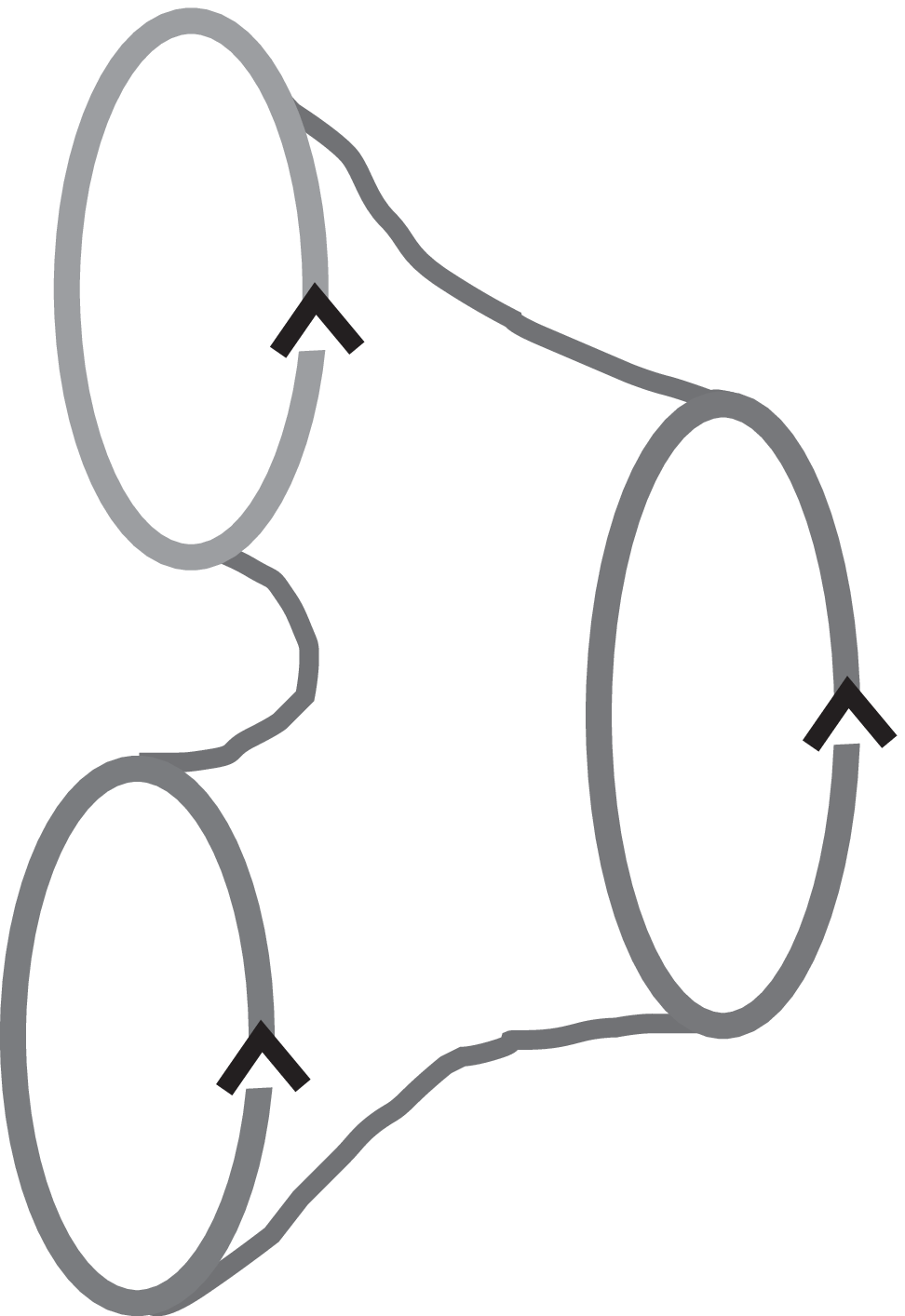}}
        \nobreak\bigskip
    {\raggedright\it \vbox{
{\bf Figure \pants.}
{\it
The 3-holed surface corresponds to the basic multiplication of the 
Frobenius algebra. 
} }}}}
    \bigskip}
 
 Let us now enlarge our geometric 
category to include open as well as closed strings. 
Now   there are ingoing/outgoing circles and 
intervals, while the morphisms are   surfaces with 
two kinds of boundaries: ingoing/outgoing boundaries
as well as ``free-boundaries,''  traced out by the 
endpoints of the in/outgoing intervals. These free boundary 
  must be labelled by ``boundary conditions'' 
which, for the moment, are merely   labels $a,b,\dots$.

\bigskip
{\vbox{{\epsfxsize=1.1in
        \nobreak
    \centerline{\epsfbox{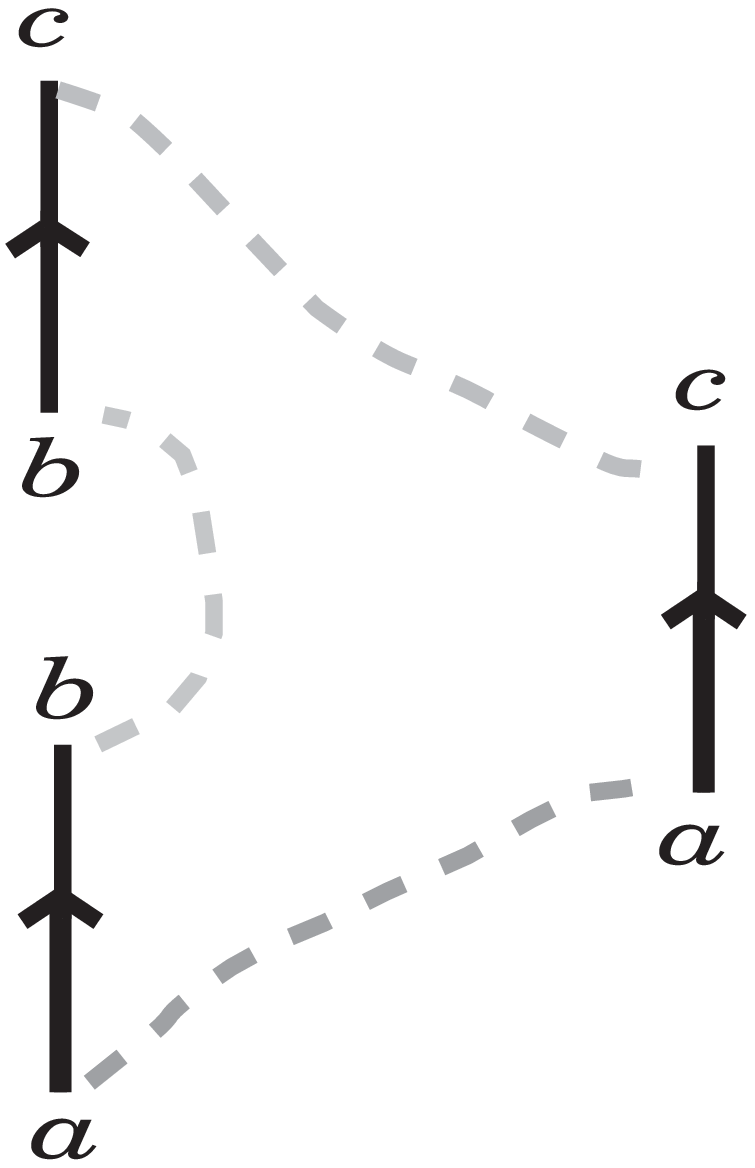}}
        \nobreak\bigskip
    {\raggedright\it \vbox{
{\bf Figure \openmult.}
{\it
Multiplication defining the nonabelian Frobenius algebra of 
open strings. } }}}}
    \bigskip}

Because we have a functor, to   any pair of boundary conditions we 
associate a vector space (``a
statespace'')  $\CH_{ab}$. Moreover, there is a coherent 
system of bilinear products
\eqn\a{
\CH_{ab}\otimes \CH_{bc} \rightarrow \CH_{ac}
}
defined by  Figure \openmult. 
This leads us to ask the key question: {\it What boundary conditions are compatible with 
$\CC$? }
``Compatibility'' means coherence with ``sewing'' or 
``gluing'' of surfaces; more precisely,   we wish to have a 
well-defined functor. Thus, just the way Fig. 1 defines an 
associative commutative algebra structure on $\CC$, Fig. 2, in 
the case $a=b=c$, defines a (not necessarily commutative) 
  algebra structure on $\CH_{aa}$. Moreover $\CH_{aa}$ is a 
Frobenius algebra. Next there are further sewing conditions 
relating the open and closed string structures. 
 Thus for example, we require that  the operators 
defined by figures \openchannel\ and \closedchannel\ be equal, a 
condition sometimes referred to as the ``Cardy condition.'' 

\bigskip
{\vbox{{\epsfxsize=1.1in
        \nobreak
    \centerline{\epsfbox{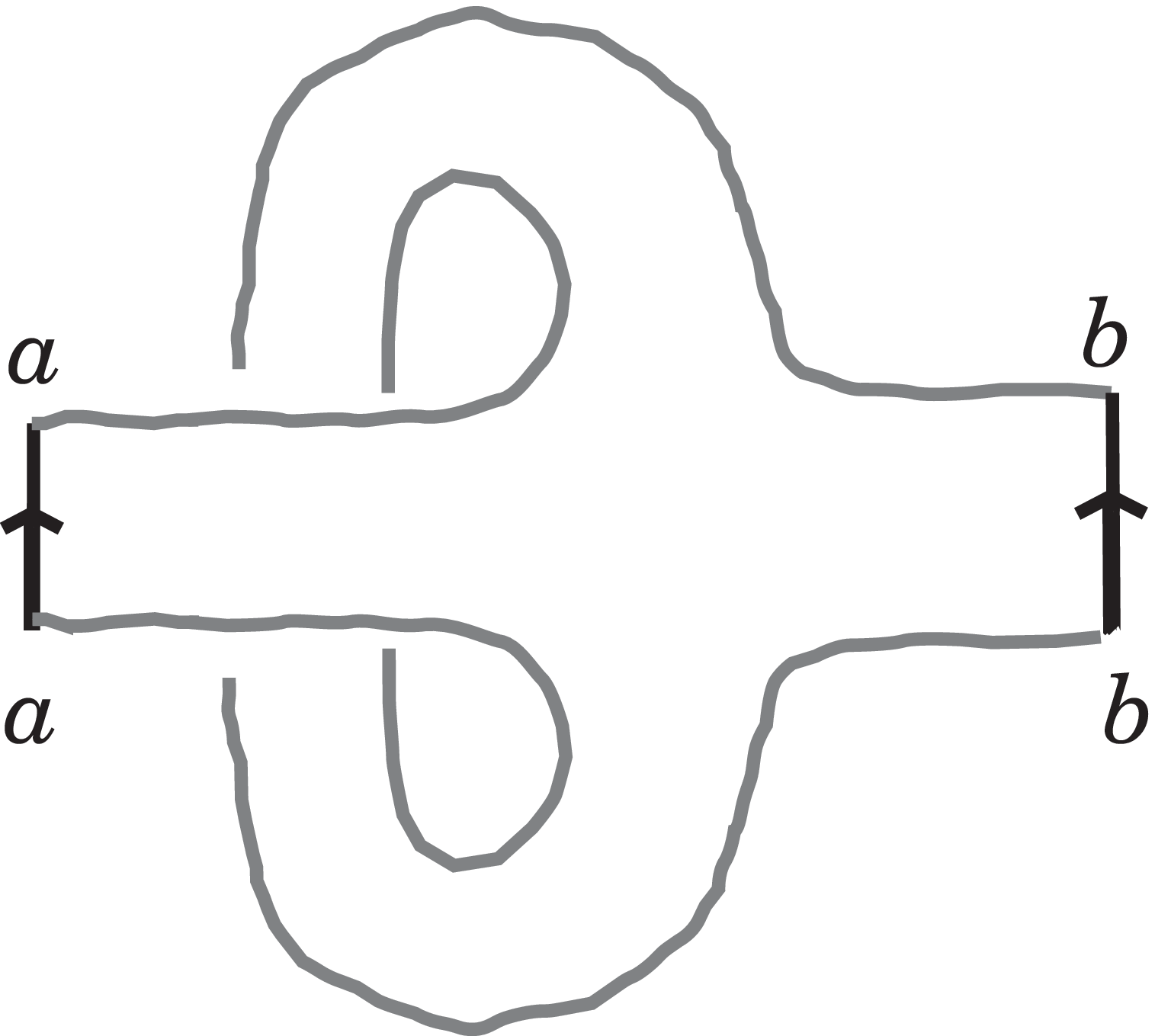}}
        \nobreak\bigskip
    {\raggedright\it \vbox{
{\bf Figure \openchannel.}
{\it
In the open string channel this surface defines a natural 
operator $\pi: \CH_{aa} \to \CH_{bb}$ on noncommutative 
Frobenius algebras. } }}}}
    \bigskip}

\bigskip
{\vbox{{\epsfxsize=1.1in
        \nobreak
    \centerline{\epsfbox{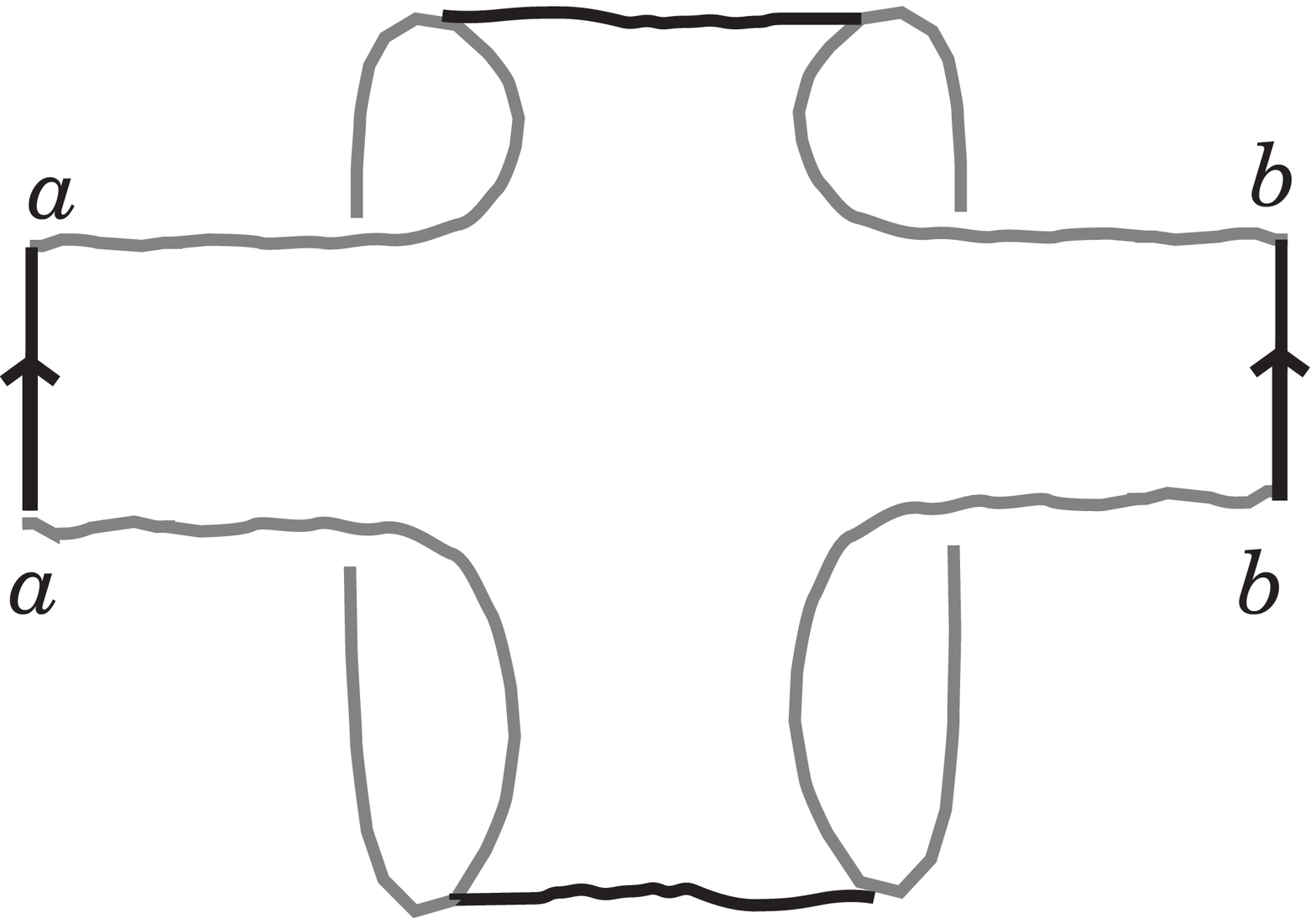}}
        \nobreak\bigskip
    {\raggedright\it \vbox{
{\bf Figure \closedchannel.}
{\it In the closed string channel this surface defines a
composition of open-closed and closed-open transitions 
 $\iota_{c-o}\iota_{o-c}: \CH_{aa} \to \CH_{bb}$ 
that factors through the center. } }}}}
    \bigskip}

As observed by Segal some time ago \segalitp, the proper interpretation 
of \a\ is that 
the boundary conditions are objects $a,b,\dots $ in an additive 
category with:
\eqn\morphsp{
\CH_{ab} = {\rm Mor}(a,b).
}
Therefore, we should ask  what the sewing constraints imply for the 
category of boundary conditions. 
This question really consists of two parts: First, coherence of sewing 
is equivalent to a certain algebraic structure on the target category. 
Once we have identified that structure we can ask for a classification 
of the examples of such structures. The first part of this question 
has been completely answered: The open/closed sewing 
conditions were first analyzed by Cardy and Lewellen \CardyTV\LewellenTB, 
and the resulting 
algebraic structure was
described in \mooresegal\LazaroiuRK. The result is   the 
following: 

\bgn
{\bf Proposition} To give an open and closed 2D oriented 
topological field theory is to give 

1. A commutative Frobenius algebra $\CC$. 

2. Frobenius algebras $\CH_{aa}$ for each boundary condition $a$. 

3. A homomorphism $\iota_a: \CC \to Z(\CH_{aa})$, where $Z(\CH_{aa})$ 
is the center, such that $\iota_a(1) = 1$, and such that, if 
$\iota^a$ is the adjoint of $\iota_a$ then 
\eqn\cardycon{
\pi_b^{~a} = \iota_b \iota^a.
}

\bigskip

Here $\pi_b^{~a}: \CH_{aa} \to \CH_{bb}$ is the morphism, determined purely in 
terms of open string data, described by Fig. \openchannel. When $\CH_{ab}$ is 
the nonzero vector space it is a Morita equivalence bimodule and $\pi_b^{~a}$
can be written as $\pi_b^{~a}(\psi) = \sum \psi^\mu \psi \psi_\mu$ where 
$\psi_\mu$ is a basis for $\CH_{ab}$ and $\psi^\mu$ is a dual basis for 
$\CH_{ba}$. More invariantly, 
\eqn\descrpi{
\theta_b(\pi_b^{~a}(\psi) \chi) = {{\Tr}}_{\CH_{ab}}\left( L(\psi) R(\chi)\right) 
}
where $\theta_b$ is the trace on $\CH_{bb}$ and $L(\psi),R(\chi)$ are the left- 
and right- representations of $\CH_{aa}$, $\CH_{bb}$ on $\CH_{ab}$, respectively.

The second step, that  of finding all examples of such structures 
was  analyzed   in \mooresegal\  in the case where 
$\CC$ is a semisimple Frobenius algebra. The 
answer turns out to be very crisp:

\bgn
{\bf Theorem 1} Let $\CC$ be 
semisimple. Then the set of isomorphism classes of objects  in the 
category of boundary conditions is
\eqn\a{ 
 K^0({\rm Spec}(\CC))= K_0(\CC) .
}
\bigskip

There are important examples of the above structure
when $\CC$ is {\it not} semisimple, such 
as the topological $A$-and $B$-twisted $\CN=2$ 
supersymmetric sigma models. As far as we are aware, 
the classification of examples for non-semisimple 
$\CC$ is an open problem.

Even in this elementary setting, there are interesting 
and nontrivial generalizations. When 
 a 2D closed topological field theory 
 has a symmetry $G$ it is possible to 
``gauge it.'' The cobordism category is enhanced by considering 
cobordisms of principal $G$-bundles.  
In this case the closed topological field theory
 corresponds to a choice of 
``Turaev algebra,'' \TuraevYF\mooresegal, 
a $G$-equivariant extension of a Frobenius algebra
which, in the semisimple case, is characterized by a ``spacetime'' 
consisting of a discrete sets of points (corresponding to the 
idempotents of the algebra), a ``dilaton,'' encoding the 
trace of the Frobenius algebra on the various idempotents, 
and a ``$B$-field.'' In this case we have 

\bgn
{\bf Theorem 2}  The isomorphism classes of objects in the 
 category   of boundary conditions 
for a $G$-equivariant open and closed theory 
 with spacetime $X $ and  ``$B$-field'' $[b]\in H^2_G(X;C^*)$
 are in 1-1 correspondence 
with the K-group of 
$G$-equivariant, $b$-twisted $K$-theory classes:
$K_{G,b}(X)$.
\bigskip

These results are, of course, very elementary. 
What I find charming about them is precisely the fact 
that   they {\it are} 
so primitive:   they rely on nothing but topological 
sewing conditions and a little algebra, and yet K-theory 
emerges ineluctably. 

A more sophisticated category-theoretic approach to the 
classification of branes in rational conformal field theories 
has been described in \refs{\FuchsQC,\FuchsYK}.

\newsec{K-theory and the renormalization group}

\subsec{Breaking conformal invariance on the boundary}

Let us now consider the much more difficult 
question of the topological classification of D-branes in 
a full  conformal field theory (CFT). This immediately 
raises the question of what   we even mean by a ``D-brane.''  
Perhaps the 
most fruitful  point of view is that 
D-branes are {\it  local  }
boundary conditions in a   2D CFT $\CC$  
which preserve conformal symmetry. While there is an 
enormous literature on the subject of D-branes, 
the  specific 
branes which have been studied  are   really a
very small subset of what is possible.

One way to approach the classification of D-branes 
  is to consider the 
space of 2D quantum field theories (QFT's), defined on 
surfaces with boundary, which are  {\it not} 
conformal, but which only break conformal invariance 
via their boundary conditions.  
Formally, there is a space $\CB$ of such boundary QFT's compatible 
with a fixed   ``bulk'' CFT, $\CC$. 
The tangent space to $\CB$ is the space of local operators on the boundary
because a local operator $\CO$ can be used to deform the action 
on a surface $\Sigma$ by: 
\eqn\b{
S_{\rm worldsheet} =S_{\rm bulk CFT}  + \int_{\p \Sigma}  ds \CO 
}
Here $ds$ is a line element. Note that in general we have introduced 
explicit metric-dependence in this term, thus breaking conformal 
invariance on the boundary.

As a simple example of what we have in mind, 
consider a massless scalar field $x^\mu:\Sigma\to \IR^n$
with action 
\eqn\bdryact{
S_{\rm worldsheet} =\int_\Sigma \p x^\mu \pb x^\mu  + \int_{\p \Sigma}  ds T\bigl(x^\mu(\tau)\bigr)
}
where  $T(x^\mu)$ is ``any function'' on $\IR^n$
and $\tau$ is a coordinate on $\p \Sigma$. 
Then the boundary interaction in \b\ can be expanded 
\eqn\c{
\CO  = T(x) + A_\mu(x){dx^\mu\over d \tau} + B_{\mu}(x) {d^2 x^\mu\over d\tau^2} + 
C_{\mu\nu}(x) {dx^\mu\over d \tau}  {dx^\nu\over d \tau}   + \cdots
}
The coefficients $T(x), A_\mu(x),\dots$ are viewed as spacetime fields on the target 
space $\IR^n$. 
\foot{G. Segal points out to me that the proper formulation of the tangent 
space to a boundary CFT would naturally use   the theory of jets. }

We expect that  $\CB$ can be given a topology 
such that renormalization group flow (see below) is a continuous evolution 
on this space. In this topology $\CB$ is 
a disconnected space. The essential idea is that 
  the {\it connected components }
of this space are classified by some kind of $K$-theory.  For example, 
if  the  conformal field theory is supersymmetric and 
has a target space interpretation in 
terms of a nonlinear $\sigma$ model,  we expect the 
components of $\CB$   to correspond to 
the K-theory of the target space $X$ 
\eqn\d{ 
\pi_0(\CB) = K(X).
}

Remarks:

\item{1.} In equation \d\ we
 are being deliberately vague about the precise form of 
K-theory (e.g. $K$, vs. $KO,KR,K_\pm$ etc.). This depends 
on a discrete set of choices one makes in formulating 
 the 2D field theory. 

\item{2.} From this point of view the importance of 
some kind of supersymmetry on the worldsheet is clear. 
As an example in the next section makes clear, the RG 
flow corresponding to taking $\CO$ to be the unit operator 
always flows to a trivial fixed point with ``no boundary.''
 Therefore,
unless the unit operator can be projected out,  there cannot 
be interesting path components in $\CB$.  In spacetime terms, we must 
cancel the ``zero momentum tachyon.''

\item{3.} One might ask what replaces \d\ when the CFT does not 
have an obvious target space interpretation. One possible answer is 
that one should define some kind of algebraic K-theory for an 
open string vertex operator algebra. There has been much recent progress 
in understanding more deeply Witten's Chern-Simons open string 
field theory (see, e.g. \refs{\BarsAG, \DouglasJM, \BarsNU,\BarsYJ,\belovlovelace}). 
 This holds out some hope that the $K$ theory of the open string 
vertex algebra could be made precise. 
In string field theory   D-branes are naturally associated to 
projection operators in a certain algebra, so the connection 
between $K$-theory and branes is again quite natural.  
See also sec. 3.5 below.

\item{4.} The space $\CB$ appeared in a proposal of Witten's 
for a background-independent string field theory \WittenQY. 
(Witten's ``other'' open string field theory.)

\item{5.} It is likely that the classification 
of superconformal boundary conditions 
in the supersymmetric Gaussian model is complete 
\refs{\friedanup,\GaberdielZQ, \JanikHB}. 
The classification is somewhat intricate and it would be 
interesting to see if it is compatible with the general 
proposal of this paper.

\subsec{ Boundary renormalization group flow  } 

One way physicists explore the path components of $\CB$ is 
via ``renormalization group (RG) flow.'' 
Since conformal invariance is broken on the boundary, 
we can ask what happens as we scale up the size of the boundary. 
This scaling defines  1-dimensional flows on $\CB$. 
These are the integral flows of a vector field $\beta$  on $\CB$
usually referred to as the ``beta function.''  
A D-brane, or conformal fixed point, corresponds to  a zero of $ \beta $.
Two D-branes which are connected by RG flow are in the same 
path component, and therefore have the same ``K-theory charge.''

Let us recall  a few facts about boundary RG flow.  For a 
good review see 
\MartinecTZ. For simplicity 
we will consider the bosonic case. 
A   boundary condition $a\in \CB$   is a zero of   $\beta$. 
At such an RG fixed point the theory is conformal, 
 and hence the Virasoro algebra acts 
on the tangent space $T_{a}\CB$. We may choose a basis of 
local operators such that $L_0 \CO_i = \Delta_i \CO_i$. 
Here $L_0$ is the scaling operator in the Virasoro algebra. 
We may then  
 choose coordinates  $\CO = \sum_i \lambda^i \CO_i$
such that, in an open neighborhood of $a\in \CB$, 
\eqn\betafun{
\beta \cong  - \sum_i (1-\Delta_i) \lambda^i {d\over d\lambda^i} 
}
Thus, as usual, perturbations by operators 
 with $\Delta_i<1$ correspond to unstable flows in the infrared (IR). 
It turns out there is an analog   of  Zamolodchikov's $c$-theorem. 
Boundary RG flow is gradient flow with respect to an ``action functional.'' 
To construct it  one introduces the natural function on $\CB$ given 
by the disk partition function. Then set 
\eqn\a{
g :=  (1+ \beta) Z_{\rm disk} .
}
Next one  introduces a metric  on $\CB$. Recalling that the local 
operators are to be identified with the tangent space we write
\eqn\met{
G(\CO_1, \CO_2) = \oint d\tau_1 d\tau_2 \sin^2({\tau_1- \tau_2\over 2}) 
\langle \CO_1(\tau_1) \CO_2(\tau_2) \rangle_{\rm disk} 
} 
Then, the   ``$g$-theorem'' states that 
\eqn\gthm{
\dot g = - \beta^i \beta^j G_{ij} 
}
The main nontrivial statement here is that $\iota(\beta)G$ is a locally 
exact one-form. 

Remarks: 

\item{1.} 
The $g$-theorem was first proposed by Affleck and Ludwig 
\AffleckTK\AffleckNG, who verified it in leading order 
in perturbation theory. An argument for the $g$-theorem, based on 
string field theory ideas,  was 
proposed in \KutasovQP\KutasovAQ. 

\item{2.} In the Zamolodchikov theorem, the c-function 
at a conformal fixed point is the value of the Virasoro central 
charge of the fixed point conformal field  theory. 
It is therefore natural to ask: ``What is the  
meaning of $g$ at a conformal fixed point?'' 
The answer is the ``boundary entropy.'' For example, when 
 the CFT $\CC$ is an RCFT with irreps  $\CH_i$, 
$i\in I$  of the 
chiral algebra, the boundary CFT's preserving the symmetry 
are labelled by $i\in I$ and the $g$-function for these 
conformal fixed points is expressed in terms of the modular 
$S$-matrix via
\eqn\bdryent{
g = {S_{0i}\over \sqrt{S_{00}}} .
}
where $0$ denotes the unit representation. 
It is notable that this can also be interpreted as a 
regularized dimension of the open string 
statespace $ \sqrt{ \dim \CH_{ii}} $. 
If the CFT is part of a string theory with a target space 
interpretation then we can go further. 
In a string theory we have gravity  and in this context 
the value of
$g$  at a conformal fixed point is 
the     brane tension, or  energy/volume of the brane
\HarveyGQ. 
 
\item{3.} In the case of $\CN=1$ worldsheet supersymmetry we 
should take instead \KutasovAQ\MarinoQC\NiarchosSI:
\eqn\susygee{
g:= Z_{\rm disk} .
}

\item{4.} In an interesting series of papers A. Connes and 
D. Kreimer have re-interpreted perturbative renormalization of 
field theory and the renormalization group in terms of the 
structure of Hopf algebras \ConnesFE\ConnesYR. We believe that the case of boundary 
RG flow in two-dimensions might be a very interesting 
setting in which to apply their ideas.

\subsec{Tachyon condensation from the worldsheet viewpoint}

Here is a simple example of the $g$ theorem.
Consider a single scalar field on the disk 
$x:D \to \IR$, where the disk $D$ has radius $r$. Then, 
\eqn\i{
Z_{\rm disk}  =  \int [dx] e^{-\int_D \p x \pb x + \oint_{\p D}T(x) } 
}
Let's just take $T(x) = t =$ constant. Then, trivially, $Z_{\rm disk}(t) = Z_{\rm disk}(0) 
e^{-2\pi r t}= Z_{\rm disk}(0)e^{-2\pi t(r)}$. 
Then 
\eqn\iii{
\beta^t:= - {\p t(r)\over \p(\log r)} = - t \quad \Rightarrow \quad 
\beta = - t{d\over dt} 
}
and an easy computation shows the metric is 
\eqn\iv{
ds^2 = e^{-t} (dt)^2.
}
The $g$-function, or action, in this case is 

\eqn\ii{
g(t) =(1+\beta) Z_{\rm disk} = (1+2\pi rt)  e^{-2\pi r t}g(0) = (1+2\pi t(r))  e^{-2\pi t(r)} g(0) 
}
At $t=0$, $Z_{\rm disk} $ is $r$-independent (hence conformally invariant) if 
we choose, say,  Neumann boundary conditions
 for $x$. Thus at $t=0$ we begin with an open/closed 
CFT consisting of  a ``D1 brane'' wrapping the target $\IR$ direction. 
 Under 
RG flow to the IR,  $t\to \infty$.  
At $t=\infty$ all boundary amplitudes are infinitely suppressed and 
``disappear.'' 
We are left with a theory only of closed strings!

Remarks: 

\item{1.} The  RG flow \ii\ is unusual in that we can give exact 
formulae.  This is 
due to its rather trivial  nature. Moreover, note that 
this boundary interaction cancels out of all 
normalized correlators. Nevertheless, we feel that the above example 
  nicely captures the essential idea. A less trivial 
example based on the boundary perturbation $\oint u X^2$ is 
analyzed in \refs{\WittenCR,\GerasimovZP,\KutasovQP}.

\item{2.} Let us return to remark 1 of section 3.1. It is 
precisely the zero-momentum tachyon (i.e. the unit operator) whose flow we wish 
to suppress in order to define a space $\CB$ with interesting path components. 

\item{3.} The  example of this section is essentially the 
``boundary string field theory'' (BSFT) interpretation of Sen's 
tachyon condensation \SenSM. In \WittenQY\  Witten introduced an 
alternative formulation of open string field theory, in 
which, (at least when ghosts decouple), the function $g$ is 
the spacetime action. This theory was further developed
by Witten and Shatashvili in 
\refs{\WittenCR,\ShatashviliKK,\LiZA,\ShatashviliPS}. Interest in the theory was 
revived by \refs{\HarveyNA,\GerasimovGA,\GerasimovZP,\KutasovQP,\KutasovAQ}. 
These papers showed, essentially 
using the above example, that the dependence of the spacetime effective 
potential on the tachyon field is 
\eqn\bspt{
V(T) \sim (T+1) e^{-T}
}
for the bosonic string and  
\eqn\iiapt{V(T) \sim e^{-T^2} 
}
for the type IIA string (on an unstable D9 brane). 
The tachyon potential is minimized by $T\to \infty$, 
and at its minimum the open strings ``disappear.''  
%

\subsec{$g$-function for the nonlinear sigma model}

 Suppose the closed CFT $\CC$ is a $\sigma$-model with 
spacetime $X$, dilaton $\Phi$, metric  $g_{\mu\nu} $
and ``gerbe connection'' $ B_{\mu\nu}$. 
A typical boundary condition involves, first of all, 
 a choice of topological $K$-homology cycle \rdouglas, that is,
an embedded subvariety $\iota: \CW \hookrightarrow X$ with 
${\rm Spin}^c$ structure
(providing appropriate Dirichlet boundary conditions for the 
open strings) together with 
a choice of (complex)  vector bundle 
\eqn\a{
E \to \CW, 
}
modulo some equivalence relations. 
We say a `` D-brane wraps $\CW$ with Chan-Paton bundle $E$.'' 

In the supersymmetric case the most  
 important boundary interaction is a choice of   a (unitary) connection $A_{\mu}$ on $E$
and a (nonabelian) section of the normal bundle. In this paper we will set the 
normal bundle scalars to zero (although they are very interesting). Thus the 
$g$-function becomes
\eqn\geesig{
g = \bigl\langle {\Tr}_E P\exp\biggl( \oint_{\p D} d\tau  A_\mu(x(\tau)) \dot x^\mu(\tau) 
+F_{\mu\nu} \psi^\mu \psi^\nu 
+ \cdots \biggr) \bigr\rangle
}
where $\psi^\mu$ are the susy partners of $x^\mu$. 
When $E$ is a line bundle $g$ can be computed for a variety of backgrounds and turns out 
to be  the Dirac-Born-Infeld (DBI) action \TseytlinDJ: 
\eqn\dbiact{
g =\int_{\CW} e^{-\Phi} \sqrt{\det_{\mu\nu} \bigl( g_{\mu\nu} +  B_{\mu\nu} + F_{\mu\nu}\bigr)} + 
\CO((DF)^2) 
}
If $E$ has rank   $ \geq 1$ to get a nice formula 
  we need to add the condition $  F_{\mu\nu} \ll 1$. In this case we have: 
\eqn\hrdbi{
g = {\rm rank}(E) \int e^{-\Phi} \sqrt{\det(g+B)} + \int_X  e^{-\Phi} {\Tr}  \bigl(F\wedge * F \bigr) + \cdots 
}

Remarks:

\item{1.} It follows from \hrdbi\ that in the long-distance limit 
the gradient flows of the  ``$g$-theorem''   generalize nicely some 
flows which appeared in the work of  Donaldson on the Hermitian-Yang-Mills equations
\donaldson.
\foot{This remark is based on discussions with M. Douglas.} 
Let $X$ be a Calabi-Yau manifold. To $X$ we associate 
an $\CN = (2,2)$ superconformal field theory $\CC$. 
The boundary interaction \geesig\ preserves 
$\CN=2$ supersymmetry iff  $F$ is of type $(1,1)$, i.e., iff $F^{2,0}=0$
\HoriCK\HoriIC. 
 RG flow preserves $\CN=2$ susy, and hence preserves 
the $(1,1)$ condition on the fieldstrength. A boundary RG fixed point is 
defined (in the $\alpha' \to 0 $ limit) 
by an Hermitian Yang-Mills connection. The RG flow is precisely the flow:  
\eqn\donflow{
{dA_\mu \over dt} = D_\nu F^{\nu}_{~~\mu}.
}
Thus, one can view the flow from a perturbation of an unstable 
bundle to a stable one as an example of tachyon condensation. 
It might be interesting to think through systematically the implications 
for tachyon condensation of 
Donaldson's results on the convergence of these flows.   

\item{2.} The tachyon  condensation from unstable $D9$ branes
(or $D9\overline{D9}$ branes) to lower dimensional 
branes involves the Atiyah-Bott-Shapiro  construction and Quillen's
superconnection in an elegant way. This has been   demonstrated 
in the context of  BSFT advocated in this section in 
\KutasovAQ\KrausNJ\TakayanagiRZ. Given the boundary data in \geesig\ one is 
naturally tempted to see a role for the ``differential K-theory'' 
described in \FreedTA. However, the nonabelian nature of the 
normal bundle scalars show that this is only part of the story. 
See \AsakawaVM\SzaboYD\AsakawaUI\   for some relevant discussions.

\subsec{The Dirac-Ramond operator and the topology of $\CB$}

Let us now make some tentative remarks on how one might try to distinguish 
different components of $\CB$. There are many indications 
that $K$-homology is a more natural framework for thinking 
about the relation of D-branes and K-theory 
\refs{\PeriwalEB,\WittenCN,\WittenNZ,\MatsuoPJ,\HarveyTE,
\MathaiIW,\AsakawaVM,
\SzaboYD,\SzaboJV}.  It was 
pointed out some time ago by Atiyah    that the 
Dirac operator defines a natural K-homology class
\atiyah. Indeed, 
abtracting the crucial properties of the Dirac operator 
leads to the notion of a Fredholm module \Connes.

Now, in string theory, the Dirac operator is generalized to 
the Dirac-Ramond operator, that is, the supersymmetry 
operator $Q$, (often denoted $G_0$) acting in the Ramond sector of a 
superconformal field theory. $Q$ is a kind of Dirac operator on 
loop space as explained in 
\refs{\WittenIM, \WittenMJ, \segalbourbaki,\stolz}. 

In the case of open strings it is still possible to define 
$Q$ in the Ramond sector, and $Q$ still has an interpretation 
as a Dirac operator on a path space. For example, suppose the 
$\CN=1$ CFT has a sigma model interpretation with closed 
string background data $g_{\mu\nu}+B_{\mu\nu}$. Suppose 
that the open string boundary conditions are $x(0)\in \CW_1$, 
$x(\pi) \in \CW_2$, where the submanifolds $\CW_i$ are equipped 
with vector bundles $E_i$ with connections $A_i$. The supersymmetry 
operator will take the form: 
\eqn\diracop{
\eqalign{
& Q = \int_0^\pi d\sigma \psi^\mu(\sigma) \Biggl( 
{\delta \over \delta x^\mu(\sigma)}  + g_{\mu\nu}(x(\sigma)){dx^\nu \over d \sigma} 
+ \left( \omega_{\mu\nu\lambda} + H_{\mu\nu\lambda} \right) \psi^\nu \psi^\lambda 
\Biggr) \cr
& + \psi^\mu(0) A_{1,\mu}(x(0)) - \psi^\mu(\pi) A_{2,\mu}(x(\pi))\cr}
}
where $\omega_{\mu\nu\lambda}$ is the Riemannian spin connection on $X$, 
and $H_{\mu\nu\lambda}$ is the fieldstrength of the $B$-field.
 Just as in 
the closed string case,  $Q$ can be understood 
more conceptually as a Dirac operator on a bundle over the path space: 
\eqn\pthsp{
\CP(\CW_1,\CW_2) = \{ x:[0,\pi] \to X \vert x(0)\in \CW_1, x(\pi)\in \CW_2\}
}
(Preservation of supersymmetry imposes further boundary conditions on 
$x$. See, for example,  \refs{\AlbertssonDV,\AlbertssonQC,\LindstromMC}
  for details. ) 
Quantization of $\psi^\mu(\sigma)$ for fixed $x^\mu(\sigma)$ 
produces a Fermionic Fock space. This space  is to be regarded as a spin 
representation of an infinite dimensional Clifford algebra. These 
Fock spaces fit together to give a  Hilbert bundle $\CS$
over $ \CP(\CW_1,\CW_2)$. 
The data $g_{\mu\nu} + B_{\mu\nu}$ induce a connection on this bundle,
as indicated in \diracop. The effect of the boundaries is merely to 
change the bundle to 
\eqn\newbundle{
\CS \to {\rm ev}_0^*(E_1)\otimes ({\rm ev}_\pi^*(E_2))^* \otimes \CS 
}
where ${\rm ev}$ is the evaluation map. The connections $A_1,A_2$ 
induce connections on \newbundle. 
In the zeromode approximation 
$Q$  becomes the Dirac 
operator on  $E_1 \otimes E_2^* \to \CW_1 \cap \CW_2$:
\eqn\quelim{
Q \to \Dsl_{E_1 \otimes E_2^*} + \cdots 
}

Now let us consider RG flow. If RG flow connects boundary 
conditions $a$ to $a'$ then the target space interpretation 
of the superconformal field theories 
$\CH_{ab}$ and $\CH_{a'b'}$ can be very different. For example, 
  tachyon annihilation can change the dimensionality of $\CW$. 
Another striking example is the decay of many D0 branes to
a single D2 brane
discussed in section 5.5  below. It follows that any formulation 
of an RG invariant involving geometrical constructions such as 
vector bundles over path space is somewhat unnatural. 
However, what {\it does} make sense throughout the renormalization 
group trajectory is the supersymmetry operator $Q$, (so long as 
we restrict attention to $\CN=1$ supersymmetry-preserving flows).
Moreover, it is physically ``obvious'' that $Q$ changes continuously 
under RG flow. This suggests that the components of $\CB$ should 
be characterized by some kind of  ``homotopy class'' of $Q$.

The conclusion of the previous paragraph immediately  
 raises the question of where the  homotopy class of $Q$ should 
take its value. We need to define a class of operators and define 
what is meant by continuous deformation within that class. 
While we do not yet have a precise proposal we can  
again turn to the zero-slope limit for guidance. 
In this limit, as we have noted, $Q \to \Dsl_{E_1 \otimes E_2^*}$, 
and $\Dsl$ defines, in a well-known way, a 
``$\theta$-summable $K$-cycle''  
for $\CA$, the $C^*$ algebra completion of $C^{\infty}(\CW_1 \cap \CW_2)$,
 acting on the Hilbert space of $L^2$ sections of $S\otimes E_1 \otimes E_2^*$ 
over $\CW_1 \cap \CW_2$. That is $[(\CH, \Dsl)] \in K^0(\CA)$ 
\Connes. What is the generalization when we do {\it not} take the 
zeroslope limit? One possibility, in the closed string case, has 
been discussed in \refs{\JaffeQZ,\JaffePZ,\ErnstBH, \Connes}. Another possibility is that 
one can define a notion of Fredholm module for vertex operator 
algebras. This has the disadvantage that it is tied to a particular 
conformal boundary condition $a$. It is possible, however, 
that the open string vertex operator algebras $\CA_{aa}$ 
for different boundary conditions $a$ are ``Morita equivalent'' 
and that the homotopy class of $Q$ defines an element of 
some $K$-theory (which remains to be defined)  ``$K^0(\CA_{aa})$.'' 
This group should be  independent of $a$ and only depend on $\CC$. 
(See section 6.4 of  \SeibergVS\ and \mooresegal\ for some discussion 
of this idea.)

Remarks: 

\item{1.} For some boundary conditions $a$ it is also possible to 
introduce a ``tachyon field.'' In this case the connection term 
$\psi^\mu(0) A_{\mu}(x(0))$ is replaced by Quillen's superconnection. 
This happens, for example, if $a$ represents a $D-\bar D$ pair with 
$\IZ_2$-graded bundle $E^+ \oplus E^-$. If the 
tachyon field $T \in {\rm End}(E^+,E^-)$ is everywhere an isomorphism 
then boundary conditions with $a$ are in the same component as the 
trivial boundary condition, essentially by the example of section 3.3.

\item{2.} 
One strong constraint on the above considerations is that the Witten 
index   
\eqn\wittenindx{
{\Tr}_{\CH_{ab}^R} (-1)^F e^{-\beta Q^2} 
}
must be a renormalization group invariant. In situations 
where we have the limit \quelim\  we can use the index theorem 
to classify, in part, the components of $\CB$. Of course, this will 
miss the torsion elements of the $K$-theory.

\newsec{$K$-theory from anomalies and instantons}

In this section we consider  the question   of understanding the 
connected components of $\CB$ in the case where there is a 
geometrical target space interpretation of the CFT. 
We will be shifting emphasis from the worldsheet to the 
target space.  
We will use an approach  based on a 
spacetime picture of branes as objects wrapping 
submanifolds of $X$ to  give an argument that (twisted) K-theory 
should classify components of $\CB$.

For concreteness, suppose our CFT is part of a background
 in type II string theory 
in a spacetime 
\eqn\a{  X_9 \times \IR,}
where the $\IR$ factor is to be thought of as time, while 
$X_9$ is compact and spin. 
Suppose moreover that spacetime is equipped with a $B$-field with 
fieldstrength $H$. This can be used to introduce a twisted 
K-group $K_H(X_9)$. 
\foot{In fact, $H$ should be refined to a 3-cocycle for integral 
cohomology. In the examples considered in detail below this 
refinement is not relevant. } 
We'll show how $K_H(X_9)$,   arises naturally in answering the question:
{\it What subvarieties of $X_9$ can a D-brane wrap? 
}
The answer involves {\it anomaly cancellation} and {\it instanton effects}, 
and leads to the slogan:  ``K-theory = anomalies modulo instantons.'' 
As an example of this viewpoint, we apply it to compute  the twisted 
$K$-theory   $K_H(SU(N))$
for $N=2,3$. We will be following the discussion of \MaldacenaXJ. 
For other discussions of the relation of twisted $K$-theory to 
Dbranes see \refs{\WittenCD,\KapustinDI,
\BouwknegtQT,\BouwknegtVU,\GawedzkiSE,\FreedTA}.
The point of view presented here has been further discussed in 
\EvslinCJ\EvslinHD.

\subsec{ What subvarieties of $X_9$ can a D-brane wrap? }
 
Since we are discussing topological restrictions and classification, we will identify 
D-brane configurations which are obtained via continuous deformation. 
Traditionally, then, we would replace the cycle $\CW$ wrapped by a D-brane by 
its homology class. This leads to the ``cohomological classification of D-branes.'' 
In the  cohomological classification of branes we follow 
two rules:

\bgn
$(A)$ Free branes
\foot{i.e., branes considered in isolation, with no other branes ending on them}
 can wrap any nontrivial homology cycle, if $\iota^*(H_{DR})$ is exact. 

\bgn
$(B)$ A brane wrapping a nontrivial homology cycle is absolutely stable. 
 
\bigskip
\bigskip

In the   K-theoretic classification of branes we have instead the modified rules:

\bgn
$(A')$ D-branes can wrap $\CW\subset X_9$ only if  $
W_3(\CW) + [H]\vert_{\CW} =0 \qquad {\rm in } \quad H^3(\CW,\IZ)$

\bgn
$(B')$ Branes wrapping homologically nontrivial $\CW$ can be unstable if, 
for some $\CW'\subset X_9$, $
PD(\CW \subset \CW') = W_3(\CW') + [H]\vert_{\CW'}.
$

Here and below, $W_3(\CW):= W_3(N\CW)$ is the Stiefel-Whitney class of the {\it normal} bundle 
of $\CW$ in $X_9$. 
\bigskip
\bigskip

We will first explain the physical reason for $(A')$ and $(B')$ and then 
explain the relation of $(A')$ and $(B')$ to twisted $K$-theory.

\bigskip
{\vbox{{\epsfxsize=1.1in
        \nobreak
    \centerline{\epsfbox{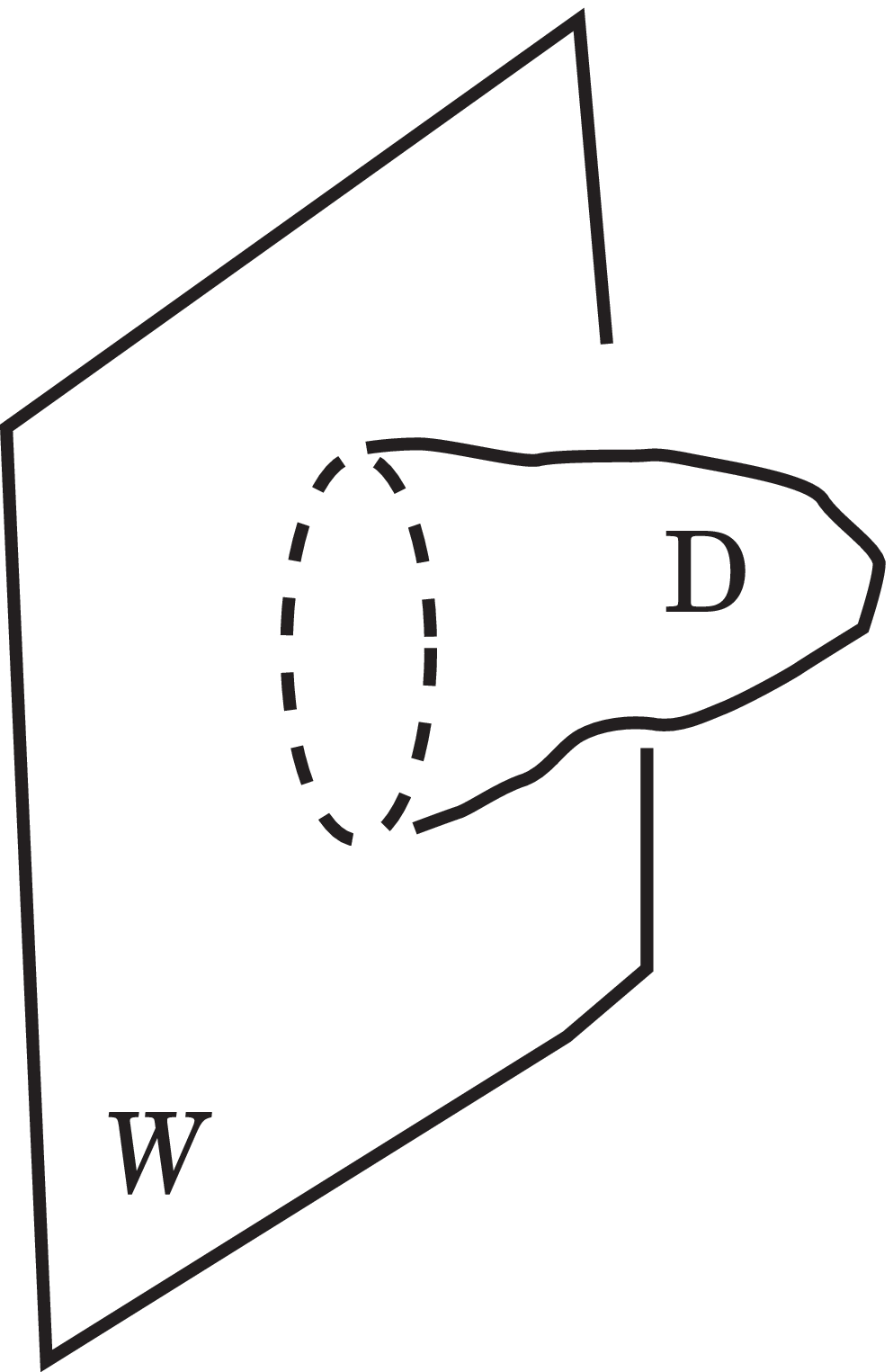}}
        \nobreak\bigskip
    {\raggedright\it \vbox{
{\bf Figure \anomcancel.}
{\it A disk string worldsheet ends on a D-brane worldvolume 
$\CW$. } }}}}
    \bigskip}

To begin, condition $(A')$ is a condition of anomaly cancellation. 
Consider a string worldsheet $D$ with boundary on a D-brane 
wrapping $\CW\times \IR$ as in Fig. \anomcancel. 
The $g$-function is, schematically 
\eqn\fgfn{
g=\int [Dx] [D\psi] 
~  e^{-\int_D \p x \bar \p x + \psi \p \psi + \cdots }~~ e^{i \int_D B} ~~ {\rm Pfaff}( \Dsl_D)
~ {\Tr}_E Pe^{i \oint_{\p D} A} 
}
The measure of the path integral must   be well-defined on the space of all maps
\eqn\maps{
\{ x: D \to X_9 \quad: \quad x(\p D) \subset {\CW} \}
}
By considering a loop of paths such that $\p D$ sweeps out a surface in $\CW$ 
it is easy to see that, at the level of the DeRham complex,
\eqn\magsource{
\iota^*(H_{DR}) = d\CF
}
must be trivialized. Heuristically   
 $\CF:=F + \iota^*(B)$, although neither $F$ nor $\iota^*(B)$ is 
separately well-defined. Note that it is the combination $\CF$ which appears in the 
$g$-function \dbiact, and 
hence {\it must} be globally well-defined on the brane worldvolume $\CW\times \IR$. 
 The equation $d\CF = \iota^*(H_{DR})$ means $H_{DR}$ is a magnetic source for $\CF$ 
on the brane worldvolume $\CW\times \IR $.

A more subtle analysis 
of global anomaly cancellation by Freed and Witten 
\FreedVC\ shows that 
\eqn\freedwitten{
\iota^*[H] + W_3(T\CW) =0
}
at the   level of integral cohomology. (See also the   discussion of
\KapustinDI.)

\bigskip
{\vbox{{\epsfxsize=1.1in
        \nobreak
    \centerline{\epsfbox{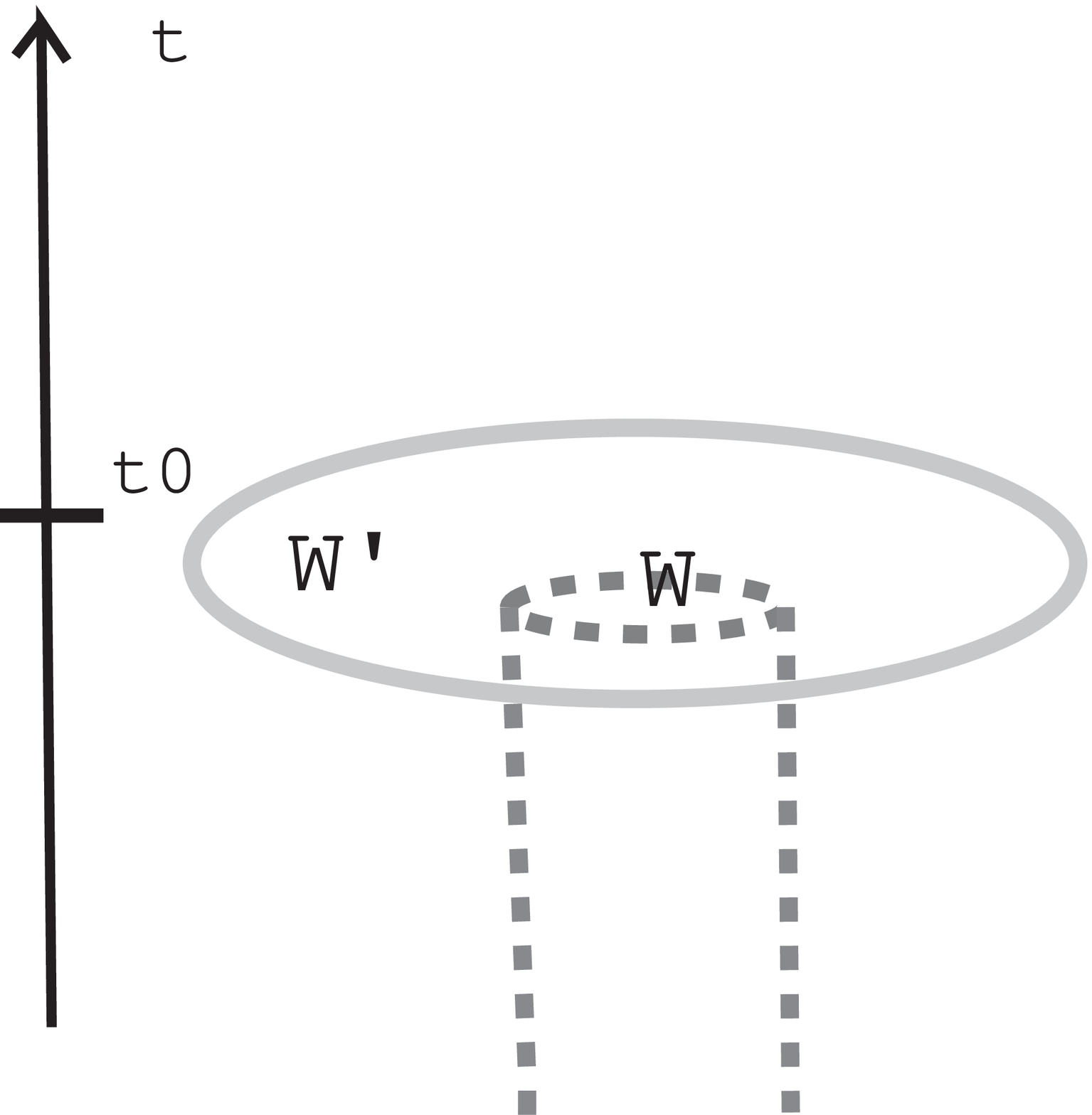}}
        \nobreak\bigskip
    {\raggedright\it \vbox{
{\bf Figure \Dinstab.}
{\it A D-brane wrapping spatial cycle $\CW$ propagates in 
time and terminates on a configuration $\CW'$ localized in 
time. This configuration of $D$-branes is anomaly-free.   } }}}}
    \bigskip}

 Let us now turn to    the stability condition  $(B')$. 
Suppose there is a cycle $\CW'\subset X_9$
on which 
\eqn\a{
W_3(\CW') + [H]\vert_{\CW'} \not=0.
}
As we have just seen, anomaly cancellation implies that we cannot wrap a D-brane on $\CW'$.  
 However, while a free  brane wrapping $\CW'$ is anomalous,   we 
 can   cancel the anomaly by 
adding a magnetic source for $\CF$. A D-brane ending on 
a codimension 3 cycle $\CW \subset \CW'$ provides such  a magentic source. 
Hence, we can construct an anomaly free configuration 
by adding a D-brane wrapping a cycle $\IR^-\times \CW$ that {\it ends} on  
   $\CW\subset \CW'$, where $\CW$  is 
such that
\eqn\a{
PD(\CW \subset \CW') = W_3(\CW') + [H]\vert_{\CW'}.
}
Here $\IR^-$ should be regarded as   a semiinfinite interval in the time-direction as in 
Fig. \Dinstab. 

  Fig. \Dinstab\   suggests a clear physical interpretation.  
 A brane wraps a 
spatial cycle $\CW$,   propagates in time, 
and terminates on a D-{\it ``instanton''} wrapping 
$\CW'$. 
This means the brane wrapping a spatial cycle $\CW$ 
can be unstable, and decays due to the configuration 
wrapping $\CW'$. 
\foot{While we use the term ``instanton'' for brevity, the 
process illustrated in figure 6  need not be nonperturbative 
in string theory. Indeed, the example of section 5.5 below 
is a process in {\it classical} string theory. The decay 
process is simply localized in the time direction. } 
  The basic mechanism is 
closely related to the ``baryon vertex'' discussed by 
Witten in the AdS/CFT correspondence \WittenXY.

\subsec{ Relation to K-theory via the Atiyah-Hirzebruch spectral sequence}

$(A')$  and $(B')$ are in fact conditions of K-theory. In order to 
understand this, let us recall the Atiyah-Hirzebruch spectral sequence
(AHSS). Let $X$ be a manifold. 
A K-theory class $x\in K^0(X)$ determines 
a system of integral cohomology classes: 
$c_i(x)\in H^{2i}(X,\IZ)$, while  
$x\in K^1(X)$ determines $\omega_{2i+1}(x)\in H^{2i+1}(X,\IZ)$. 
Let us ask the converse. 
  Given a system of cohomology classes, $(\omega_1, \omega_3, \dots)$
does there exist an $x\in K^1(X)$ ? 
The AHSS is a successive approximation scheme: $E_1^*, E_3^*, E_5^*,\dots$
 for 
describing when such a system of cohomology classes 
$(\omega_1, \omega_3, \dots)$ arises from a K-theory class.

In order to relate the AHSS   to D-branes we regard 
$PD(\omega_k)$ in $X$  as the  spatial cycle of a (potentially 
unstable)   brane of spatial dimension $\dim X - k$. 
In this way, a system $(\omega_1, \omega_3, \dots)$ determines a collection of branes, 
and hence the AHSS helps us decide which subvarieties of $X$ can be wrapped. 
Now let us look at the AHSS in more detail: 

The first approximation is the cohomological classification of  D-branes:  
\eqn\ahssi{
\eqalign{ 
K^0(X) \sim   E^0_1(X) & :=   H^{\rm even}(X,\IZ) \cr
K^1(X) \sim  E^1_1(X) & :=  H^{ \rm odd}(X,\IZ) \cr}
}
The  
first nontrivial approximation is
\eqn\ahssii{
\eqalign{ 
K^0(X) \sim E^0_3(X) &  := \bigl( {\rm Ker} ~d_3\vert_{H^{\rm even} }\bigr) / \bigl({\rm Im} ~d_3\vert_{H^{\rm odd}}\bigr)\cr
K^1(X) \sim E^1_3(X)  & := \bigl( {\rm Ker} ~d_3\vert_{H^{\rm odd} }\bigr) / \bigl({\rm Im} ~d_3\vert_{H^{\rm even}}\bigr)\cr
}
}
with 
\eqn\ahssiii{
d_3(a) := Sq^3(a) + [H]\smile a.
}

Let us pause to define $Sq^3(a)$. Let us suppose, for simplicity, that the Poincar\'e dual 
$PD(a)$ can be represented by a manifold $\CW$ and let    $\iota: \CW \hookrightarrow X_9$ be the 
inclusion. Then we let 
\eqn\sqthree{
Sq^3(a) = \iota_*(W_3(\CW) )  
}
where $\iota_*$ is a composition of three operations: 
first take the Poincar\'e dual of $W_3(\CW)$ within $\CW$, then push forward the 
homology cycle, and then take the Poincar\'e dual in $X_9$. Equivalently, 
regard $a$ as a class compactly supported in a tubular neighborhood of $\CW$ and 
consider the class $W_3(\CW) \smile a$ where $W_3(\CW)$ is pulled back to the tubular 
neighborhood.

Returning to the AHSS, in  general one must continue the approximation scheme. This is  
true, for example, when computing the twisted K-theory of    $SU(N)$ for $N\geq 3$.

 Now, let us  interpret the procedure of taking $d_3$ cohomology in physical 
terms.  To interpret   ${\rm Ker }~ d_3$  
note that from \sqthree\ it follows that  
\eqn\a{d_3(a)=0 \quad \Leftrightarrow \quad 
\biggl( W_3(\CW) + [H] \biggr) \smile a = 0 .
}
Recall that 
global anomaly 
cancellation for a D-brane wrapping $\CW$ implies 
\eqn\a{
 W_3(\CW) + [H]\vert_{\CW} =0 
}
and this in turn implies $d_3(a)=0$. 
Thus, the 
physical condition $(A')$  implies   $PD(\CW) \in {\rm Ker} d_3$. 

 Next, let us interpret   the quotient by the image of $d_3$
in \ahssii. 
Suppose $a=d_3(a') = (Sq^3 + [H])(a')$. 
Then, choose representatives
\eqn\a{
PD(a) = \CW \qquad  PD(a') = \CW' , 
}
where
$\CW$ is codimension 3 in $\CW'$. A $D$-brane terminating on 
$\CW$ can be the magnetic source for the D-brane 
gauge field on $\CW'$ and 
\eqn\a{
PD\bigl( \CW \hookrightarrow \CW' \bigr) 
= W_3(\CW') + [H]\vert_{\CW'} ~~ \Rightarrow ~~a = d_3(a') 
}
Therefore, the physical process of D-instanton induced brane instability 
implies one should take the quotient by the image of $d_3$
\DiaconescuWY\MaldacenaXJ. (In fact, conditions $(A'),(B')$ contain more 
information than $d_3$.)

\subsec{Examples: Twisted K-groups of $SU(N)$} 

As an illustration of the above point of view let us 
consider the twisted K-groups of $SU(2)$ and $SU(3)$.

Consider first  $K_H(SU(2))$. Then $H=k \omega$ where    
$\omega$ generates $H^3(SU(2);\IZ)$. 
In the cohomological model of branes we have  
$H^{\rm even}(SU(2))= H^0 \cong H_3= \IZ $, 
corresponding to  ``D3-branes'' (or D2-instantons) 
while  
$H^{\rm odd}(SU(2))= H^3 \cong H_0 = \IZ $ 
corresponding to ``D0-branes.'' 
Now condition 
$(A')$  shows that we can only have D0 branes. 
Indeed, D2 instantons wrapping $SU(2)= S^3$ violate D0-brane 
charge by $k$ units as in Fig. \Dviol. For this reason if we 
 take $SU(2)$ as the cycle $\CW'$ in condition 
$(B')$ then it follows that   a system with $k$ D0 branes is 
in the same connected component of $\CB$ as a system with 
no D0 branes at all. In     section 5.5 we will explain in 
more detail how this can be.

\bigskip
{\vbox{{\epsfxsize=1.1in
        \nobreak
    \centerline{\epsfbox{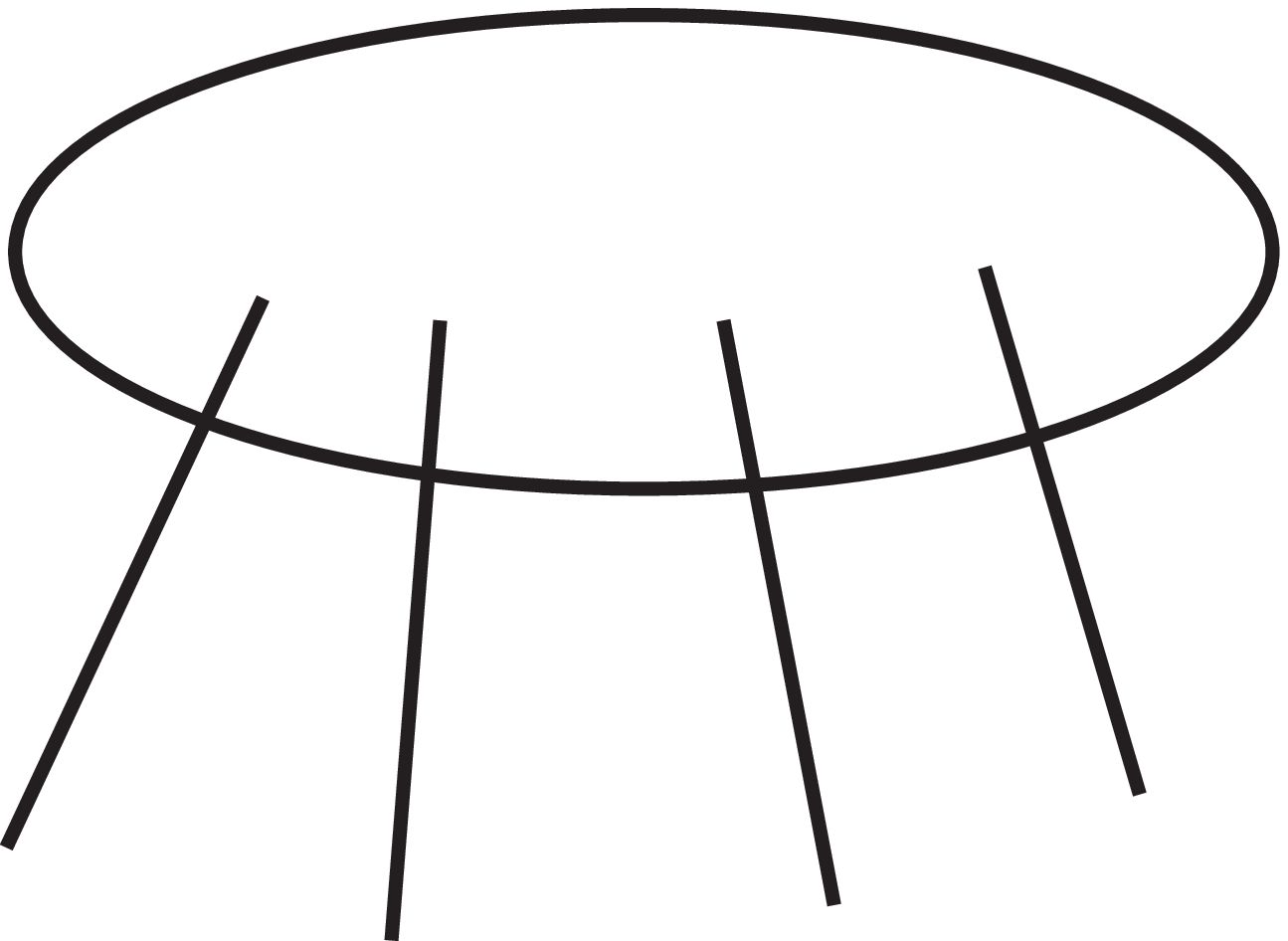}}
        \nobreak\bigskip
    {\raggedright\it \vbox{
{\bf Figure \Dviol.}
{\it  $k$ D0 branes terminate on a wrapped D2-brane 
instanton in the $SU(2)$ level $k$ theory.   } }}}}
    \bigskip}

In this way we conclude that 
\eqn\a{
\eqalign{
K^0_H(SU(2)) & = 0 \cr
K^1_H(SU(2)) & = \IZ/k\IZ\cr}
}
as is indeed easily confirmed by rigorous mathematical arguments.

Let us now consider    $K_H(SU(3))$. Here the AHSS is not 
powerful enough to determine the K-group. However, it is  
 important to bear in mind that 
the physical conditions $A',B'$ contain 
more information, and are stronger, than 
the $d_3$-cohomology.  Once again we take  $H=k \omega$, 
where $\omega$ generates $H^3(SU(3);\IZ)$.  

In the cohomological model we have 
$H^{\rm odd} \cong H_3 \oplus H_5$. 
Now, 
3-branes cannot wrap $SU(2)\subset SU(3)$ since $[H]_{DR}\not=0$. 
But 5-branes can wrap the cycle $M_5\subset SU(3)$, where 
$M_5$ is 
Poincar\'e dual to $\omega$.  
Now, 
\eqn\steensqr{
\int_{SU(3)} \omega Sq^2 \omega = 1
}
and hence 
\eqn\wthre{\iota^*(\omega) = W_3(M_5)}
is nonzero. (In fact, it turns out that the cycle $M_5$ 
can be represented by the space of symmetric 
$SU(3)$ matrices. This space is diffeomorphic to
 $ SU(3)/SO(3)$ and is a simple example of 
a non-Spin$^c$ manifold.) 

It follows from  \wthre\ that 
if $M_5$ is wrapped $r$ times, anomaly cancellation implies
\eqn\a{
r(k+1) W_3 = 0 
}
The D-brane instantons relevant to condition $(B')$ are 
just the $D2$-branes wrapping $SU(2)$. 
We thus conclude that 
\eqn\rstl{
K^1_{H=k\omega}(SU(3))  = \cases{
 \IZ/k\IZ & $k$ odd  \cr
  2\IZ/k\IZ & $k$   even \cr}
}
 
Let us now turn to the even-dimensional branes, 
  $H^{\rm even} \cong H_0 \oplus H_8$. 
8-branes are anomalous because $[H]_{DR}\not=0$, but 
$0$-branes are anomaly-free.

\bigskip
{\vbox{{\epsfxsize=1.1in
        \nobreak
    \centerline{\epsfbox{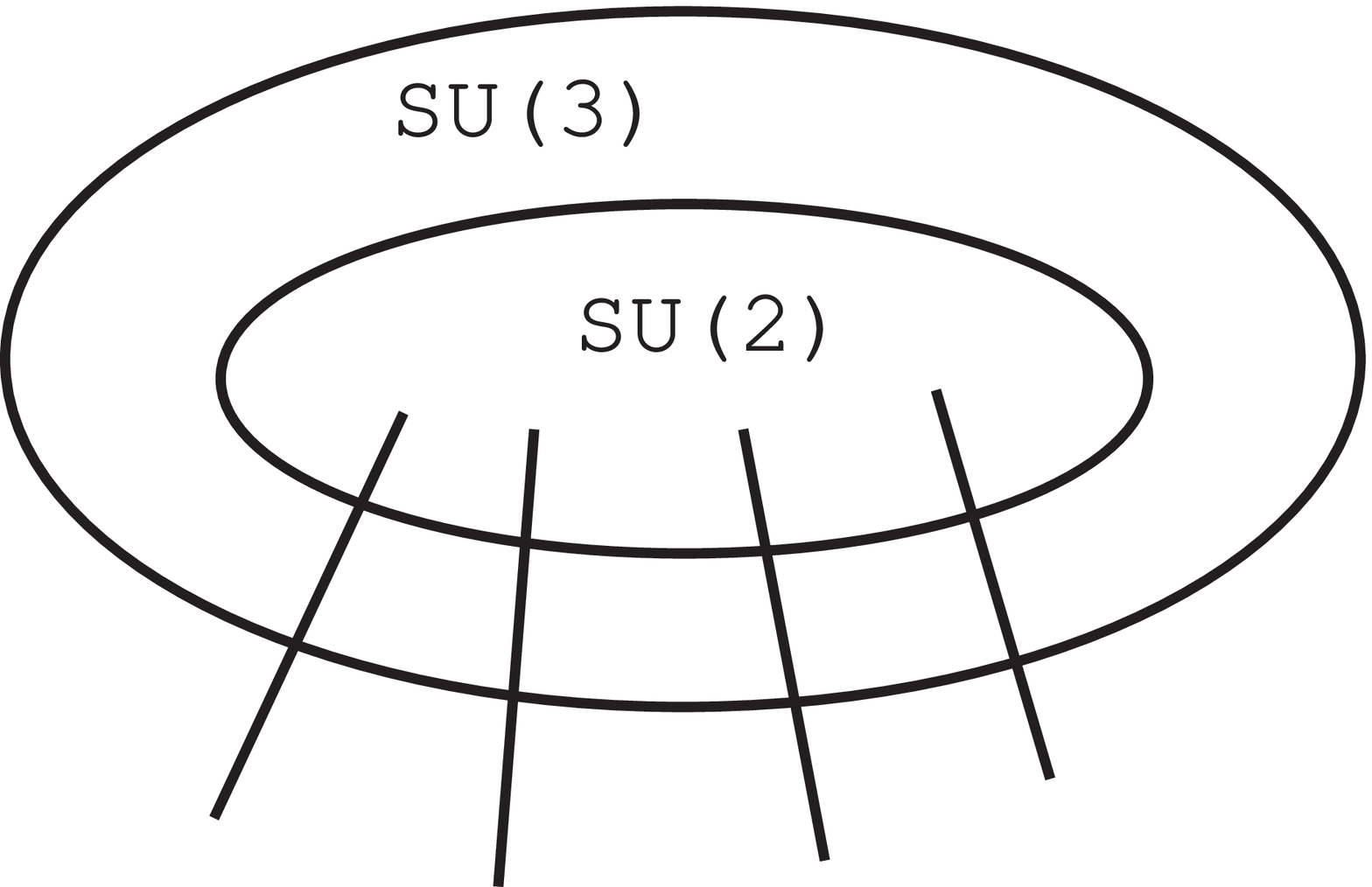}}
        \nobreak\bigskip
    {\raggedright\it \vbox{
{\bf Figure \Dviolii.}
{\it  $k$ D0 branes terminate on a wrapped D2-brane 
instanton in an $SU(2)$ subgroup of $SU(3)$   } }}}}
    \bigskip}

\bigskip
{\vbox{{\epsfxsize=1.1in
        \nobreak
    \centerline{\epsfbox{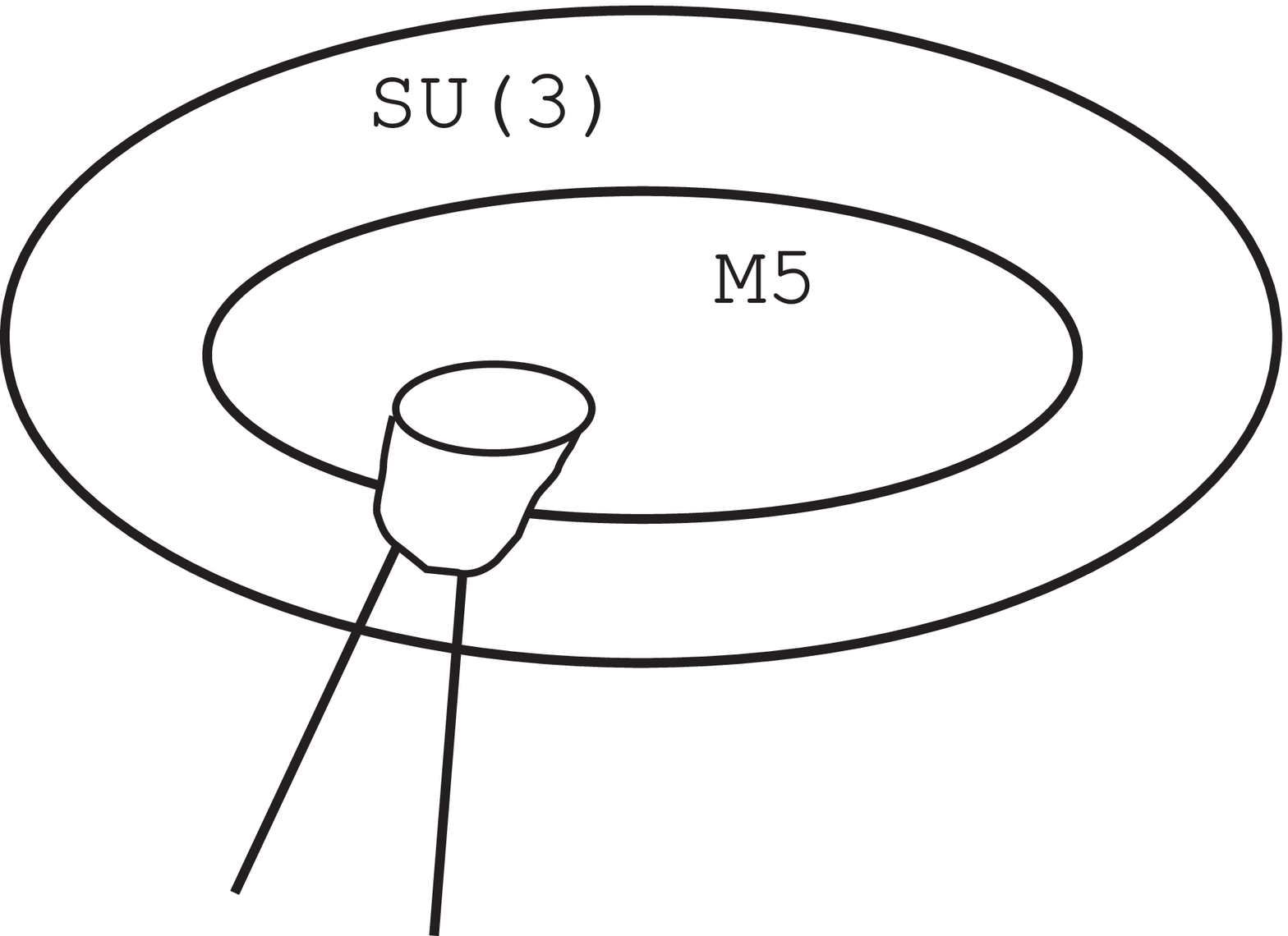}}
        \nobreak\bigskip
    {\raggedright\it \vbox{
{\bf Figure \Dvioliii.}
{\it  When $k$ is even $\half k$ D0 branes can terminate on a
hemisphere of $SU(2)$ which terminates on a generator of 
$H_5(SU(3),\IZ)$.   } }}}}
    \bigskip}

Now, D0-brane charge is not conserved
because of the standard process of Fig. \Dviolii. 
There is, however, a more 
subtle instanton, illustrated in \Dvioliii, 
in which a 3-chain ends on a nontrivial element in  
 $H_2(M_5;\IZ)\cong \IZ_2$.  This instanton 
violates D0 charge by $\half k$ units, when $k$ is 
even. 
In this way we conclude that 
\eqn\a{
K^0_{H}(SU(3)) = \cases{
   \IZ/k\IZ & $k$  odd \cr
   \IZ/{k\over 2}\IZ  & $k$   even \cr}
}

One could probably extend the above procedure to compute the 
twisted K-theory of higher rank groups
using (at least for $SU(N)$),  Steenrod's cell-decomposition, 
but this has not been done. In part inspired by the 
above results (and the result for D0 charge quantization 
explained in the next section) M. Hopkins computed the 
twisted $K$-homology of $SU(N)$ rigorously.  
He finds that,   for $H=k\omega$:   
\eqn\hop{
K_{H,*}(SU(N)) = 
(Z/d_{k,N} Z)\otimes \Lambda_Z[w_5, \dots, w_{2N-1}]
}
where
\eqn\dzero{
d_{k,N}  = {\rm gcd}\bigl[{k\choose 1}, {k\choose 2},
\dots, {k \choose N-1}\bigr].
}
We find perfect agreement   for $G=SU(2),SU(3)$ above. 

It is interesting to compare \hop\  with  
\eqn\a{
H_*(SU(N)) = \Lambda_Z[w_3, w_5, \dots, w_{2N-1}].
}
Recall that $SU(N) \sim S^3 \times S^5 \times \cdots S^{2N-1}$, rationally. 
Evidentally, the topologically distinct D-branes can be pictured as 
wrapping different cycles in $SU(N)$, subject to certain decay 
processes. In the next section we will return to the worldsheet 
RG point of view    to explain the 
most important of these decay processes.  We will also
give a simple physical argument (which in fact predated 
Hopkins' computation) for why the group of charges should be 
torsion of order $d_{k,N}$.

\newsec{The example of branes in $SU(N)$ WZW models}

In this section we will use the theory of 
``symmetry-preserving branes'' to determine the order  
$d_{k,N}$ of the D0 charge group for $ SU(N) $ level $k$ 
WZW model. Different versions of the argument are 
given in \refs{  \AlekseevJX,\FredenhagenEI,\MaldacenaXJ,\StanciuVW}. 
For reviews with further details on the material of this section see 
\SchomerusDC\SchweigertIX\GawedzkiYE\ and references therein.

Let us summarize the strategy of the argument here:  
 
\item{1.} We define  the ``elementary'' or ``singly-wrapped''
 symmetry-preserving boundary conditions algebraically using the formalism 
of boundary conformal field theory. These boundary 
conditions are labelled by
the unitary irreps $\lambda \in P^+_k$ of the centrally extended loop group. 

\item{2.} We   give a semiclassical picture of these boundary 
conditions as branes wrapping special regular conjugacy classes with 
a nontrivial Chan-Paton line bundle. See equations (5.14) and (5.24)  below. 
 
\item{3.} We then discuss how it is that ``multiply-wrapped'' 
symmetry preserving branes can lie in components of $\CB$ 
corresponding to certain singly-wrapped branes. For example, 
a ``stack of $L$ D0  branes'' can be continuously connected by 
RG flow to a symmetry-preserving brane labelled by $\lambda$, 
provided the number of D0 branes $L$ is equal to the dimension  
 $d(\lambda)$  of the 
representation $\lambda$ of the group $G$.

\item{4.} This implies that the symmetry-preserving brane 
$\lambda$ has, in some sense, D0 charge $L$. On the other hand, 
as we have seen in the previous section, the D0 charge must 
be finite and cyclic. Thus the D0 charge is $d(\lambda) {\rm mod} d_{k,N}$, 
for some integer $d_{k,N}$. 

\item{5.} Finally we note that symmetry-preserving 
branes for different values of $\lambda$ can sometimes be
  related by a rigid rotation continuously connected to 1. 
Such branes  are obviously in the same component of $\CB$, 
and this suffices to determine the order
$d_{k,N}$ of the torsion group.

\subsec{ WZW Model for $G=SU(N)$}

Let us set our notation. The WZW field $g: \Sigma \to G$ has action 
\eqn\wzw{
S = {k\over 8 \pi} \int_{\Sigma}   {\Tr}_N [
(g^{-1}\p g) (g^{-1}\pb  g)] + {2 \pi k } \int \omega
}
where the trace is in the fundamental representation. The  
 target space     $G=SU(N)$ has a  metric 
\eqn\a{
 ds^2 = -{k \over 2} {\Tr}_N (g^{-1}dg\otimes g^{-1} dg)
}
and a   ``$B$-field'' with fieldstrength 
\eqn\wzwb{
H = k \omega, \qquad   \quad \omega:= -{1\over 24 \pi^2}{\Tr}(g^{-1} dg)^3
}
where $[\omega]$ generates $H^3(G;Z) \cong Z$. 
The CFT state space is  
\eqn\a{
\CH^{\rm closed} \cong \oplus_{\rm P^+_k} \CH_\lambda \otimes \tilde \CH_{\lambda^*}
}
where $\CH_{\lambda},  \tilde \CH_{\lambda^*}$ are the 
left- and right-moving unitary irreps of the   loop group $\widetilde{LG}_k$, as
described in \pressleysegal.  

Amongst the set of conformal boundary conditions (i.e. branes) there is a 
distinguished set of   ``symmetry-preserving boundary 
conditions'' leaving the diagonal sum of left and right-moving 
currents $J + \tilde J$ unbroken. See     \SchomerusDC\SchweigertIX\ 
for more details. Since there is an unbroken affine symmetry 
the open string 
morphism spaces $\CH_{ab}^{\rm open}$ 
are themselves representations of $\widetilde{LG}_k$. 
Accordingly, these are objects in the category of boundary 
conditions labelled by $\lambda\in P^+_k$.  The decomposition 
of the morphism spaces as irreps of $\widetilde{LG}_k$ is given by 
\eqn\wzwopen{
\CH_{\lambda_1, \lambda_2}^{\rm open}
 = \oplus_{ \lambda_3\in P^+_k} N^{  \lambda_3}_{\lambda_1, \lambda_2} \CH_{  \lambda_3} 
}
where $N^{  \lambda_3}_{\lambda_1, 
\lambda_2}$ are the fusion coefficients.

The most efficient way to establish \wzwopen\ is via 
the ``boundary state formalism.'' In the 2D topological 
field theory of section 2, the boundary state associated to 
boundary condition $a$ is defined to be $\iota^{a}(1_a)$ where 
$1_a$ is the unit in the open string algebra $\CH_{aa}$. 
This is an element of the closed string algebra $\CC$ which 
``creates'' a free boundary with boundary condition $a$. 
Similarly, in   boundary CFT, to every conformal boundary 
condition $a$ one associates a 
corresponding ``boundary state'' 
\eqn\a{
\vert B(a)\rara \in \CH^{\rm closed}.
} 
For the symmetry-preserving WZW boundary conditions 
the corresponding   boundary state is given by 
the Cardy formula: 
\eqn\spbs{ 
\vert B(\lambda) \rara= \sum_{\lambda'\in P^+_k} {S_{\lambda}^{~\lambda'} \over \sqrt{
S_{0}^{\lambda'} }} ~ {\bf 1}_{\CH_{\lambda'}} \in \CH^{\rm closed} 
}
where $S_{\lambda}^{~\lambda'}$ is the modular $S$-matrix, 
$0$ denotes the basic representation,  and  
 we think of the closed string statespace as: 
\eqn\spbsi{
\CH^{\rm closed} 
\cong \oplus_{ P^+_k} \CH_\lambda \otimes \tilde \CH_{\lambda^*}
  \cong 
 \oplus_{ P^+_k}   {\rm Hom}(\CH_{ \lambda} , \CH_{\lambda} )
}
Applying the Cardy condition to \spbs\ we get \wzwopen. 

Since the disk partition function is the overlap of the 
ground state with the boundary state, 
$Z_{\rm disk} = \langle 0 \vert B(\lambda)\rangle\rangle$, 
the $g$-function   for these conformal fixed points follows 
immediately from \spbs:  
\eqn\bdryent{
g(\lambda) = S_{\lambda,0}/\sqrt{S_{00}} .
}
 
Finally, as we noted before, it is important to introduce worldsheet
supersymmetry in order to have any stable branes at all. 
It suffices to introduce $\CN=1$ supersymmetry, although when 
 embedded in a 
type II string background 
the full background can have $\CN=2$ supersymmetry. 
It is also important to have a well-defined action by $(-1)^F$ on the 
conformal field theory. This distinguishes the cases where the 
rank of $G$ is odd and even. When the rank is odd we can always add 
an $\CN=1$ Feigin-Fuks superfield, as indeed is quite natural 
when building a type II string background.

\subsec{Geometrical interpretation of the symmetry-preserving branes}

We would like to discuss the {\it geometrical} interpretation 
of the symmetry-preserving boundary condition labelled by $\lambda$. 
That is, we would like some semiclassical picture of 
the brane as an extended object in the group manifold.
In this section we explain how that is derived.

Let us first recall how the geometry of the compact target space 
is recovered in the 
WZW model. 
In the WZW model the metric is proportional to $k$, so  
the path integral measure has weight factor
\eqn\a{\sim e^{ - k S}}
Thus, we expect  semiclassical pictures to emerge
in the limit $k\to \infty$.
 In this limit the vertex operator algebra ``degenerates'' to become 
the algebra of functions on the group $G$. For example, 
CFT correlators become integrals over the group manifold: 
\eqn\a{
\langle \hat F_1(g(z_1,\bar z_1)) \cdots  \hat F_n(g(z_n,\bar z_n))\rangle 
\to \int_G d\mu(g) F_1(g)\cdots F_n(g)
}
On the left hand side $\hat F_i$ are suitable vertex operators
of  dimension $\sim 1/k$.  On the right hand 
side, $F_i$ are corresponding $L^2$ functions on $G$. 
Roughly speaking, the CFT statespace degenerates as 
\eqn\statlim{
\CH^{\rm closed} 
\cong \oplus_{\rm P^+_k} \CH_\lambda \otimes \tilde \CH_{\lambda^*}
\rightarrow L^2(G)\otimes \CH^{\rm string} 
}
where $L^2(G)$ is the limit of the primary fields and 
$\CH^{\rm string}$ contains the ``oscillator excitations.'' 
In this limit the boundary state degenerates: 
\eqn\btlim{
\vert B(\lambda)\rara
\to B_\lambda  + \cdots  
}
where $B_\lambda\in L^2(G)$ and becomes a distribution in 
the $k \to \infty$ limit. While \statlim\ is 
clearly heuristic, \btlim\ has a well-defined meaning 
because the overlaps of $\vert B(\lambda) \rara$ with 
primary fields of dimension $\sim 1/k$ 
have well-defined limits.

\bigskip
{\vbox{{\epsfxsize=1.1in
        \nobreak
    \centerline{\epsfbox{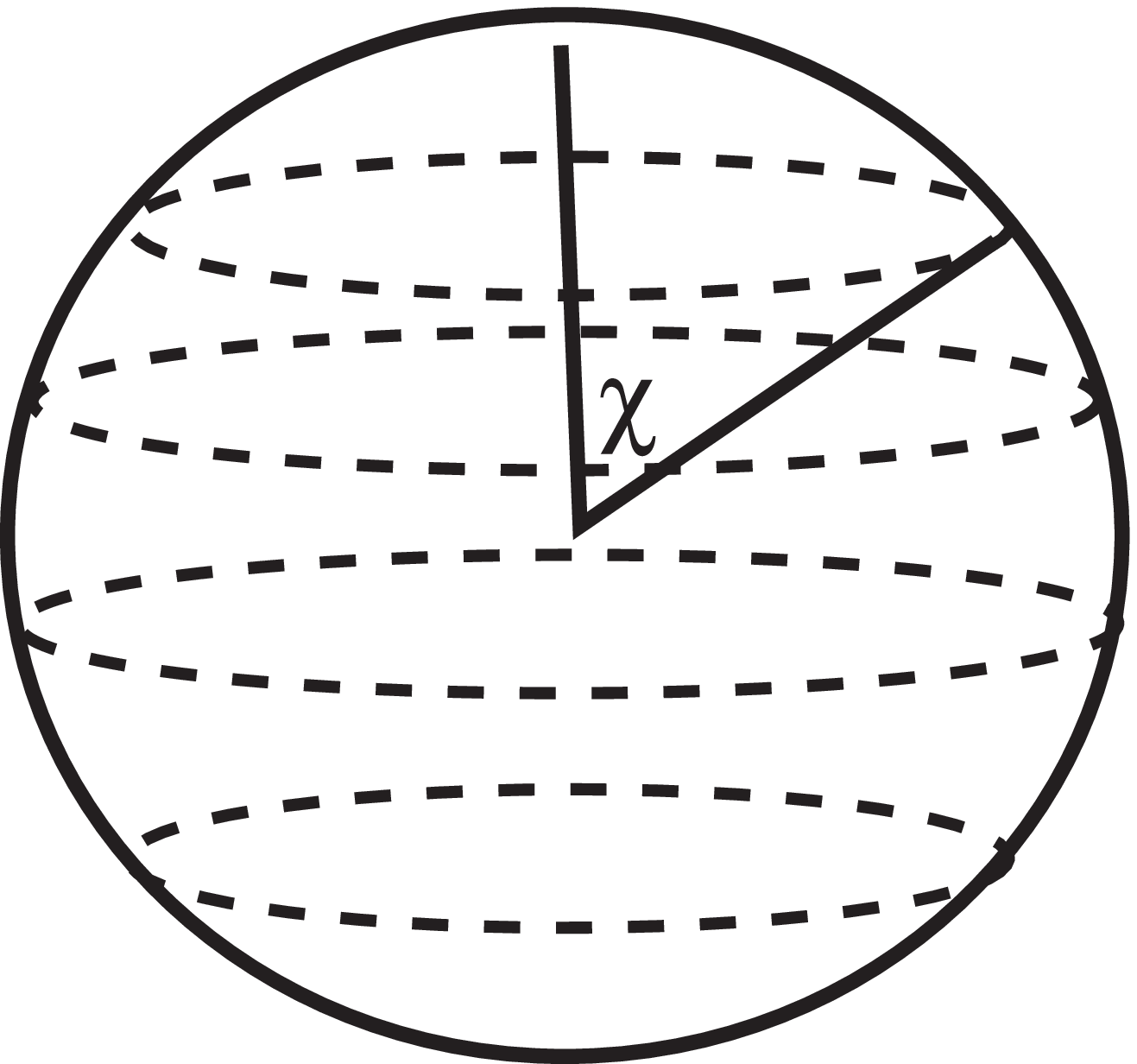}}
        \nobreak\bigskip
    {\raggedright\it \vbox{
{\bf Figure \symmbrane.}
{\it   Distinguished conjugacy classes in $SU(2)$. These are the semiclassical 
worldvolumes of the symmetry-preserving branes.   } }}}}
    \bigskip}

Using  equation \spbs, the formulae 
for the modular $S$-matrix, and the Peter-Weyl theorem one finds 
that the  function $B_\lambda$ is concentrated 
on the regular conjugacy class
\eqn\regcc{\CO_{\lambda,k}:= \biggl[ \exp\bigl(2\pi i {\lambda + \rho \over k + h}  \bigr)\biggr]
}
leading to the semiclassical picture of branes in Fig. \symmbrane\  above. 
Here $\rho$ is the Weyl vector and $h$ is the dual Coxeter number. 
(As usual, replace $k \to k-h, \lambda\in P^+_{k-h}$ for the 
supersymmetric case.) 
See \FelderKA\MaldacenaKY\MaldacenaXJ\  for more details.

\bgn 
Remarks:

\item{1.}   Since $k$ is the semiclassical expansion parameter 
we only expect to be able to localize the branes to within a  
 length-scale $\ell_{\rm string} \sim 1/\sqrt{k}$
when using closed string vertex operators \KlebanovNI. Let 
$\liet$ be the  
the Lie algebra of the maximal torus, and let 
$\chi\in \liet$ 
 parametrize conjugacy classes. Then  
the metric $ds^2 \sim k (d\chi)^2$ and hence vertex operators 
can only ``resolve'' angles $\delta \chi\geq {1\over \sqrt{k}}$.  
This uncertainty encompasses many different conjugacy classes \regcc. 
Nevertheless, the semiclassical geometrical pictures give exact 
results for many important physical quantities. 
The reason for this is that the relevant exact CFT results are polynomials 
in $1/k$, and hence can be exactly computed in a semiclassical 
expansion. 

\item{2.} The basic representation $\lambda=0$ gives the 
``smallest'' brane. We will refer to this as a ``D0-brane.'' 
In a IIA string compactification built with  the WZW model 
this state is used to construct a D0 brane. Note, however, 
that in this description it is {\it not} pointlike, but rather 
has a size of order the string length $\sim 1/\sqrt{k}$. 

\subsec{Using CFT to measure the distance between branes}

To lend further support to the geometrical picture advocated 
above, let us show that 
D-branes can be rotated in the group, and that the distance 
between them can then be measured using  CFT techniques.

First, we explain how to ``rotate'' D-branes.  
$G_L \times G_R$ acts on $\CH^{\rm closed}$, 
and therefore we can consider the boundary state 
\eqn\nwbry{
 g_L  g_R\vert B(\lambda) \rara.
}
In the $k\to \infty$ limit this
state 
has a limit similar to \btlim. In particular, it   is  supported 
on the   subset 
\eqn\a{
g_L \CO_{ \lambda,k } g_R \subset G.
}

\bigskip
{\vbox{{\epsfxsize=1.1in
        \nobreak
    \centerline{\epsfbox{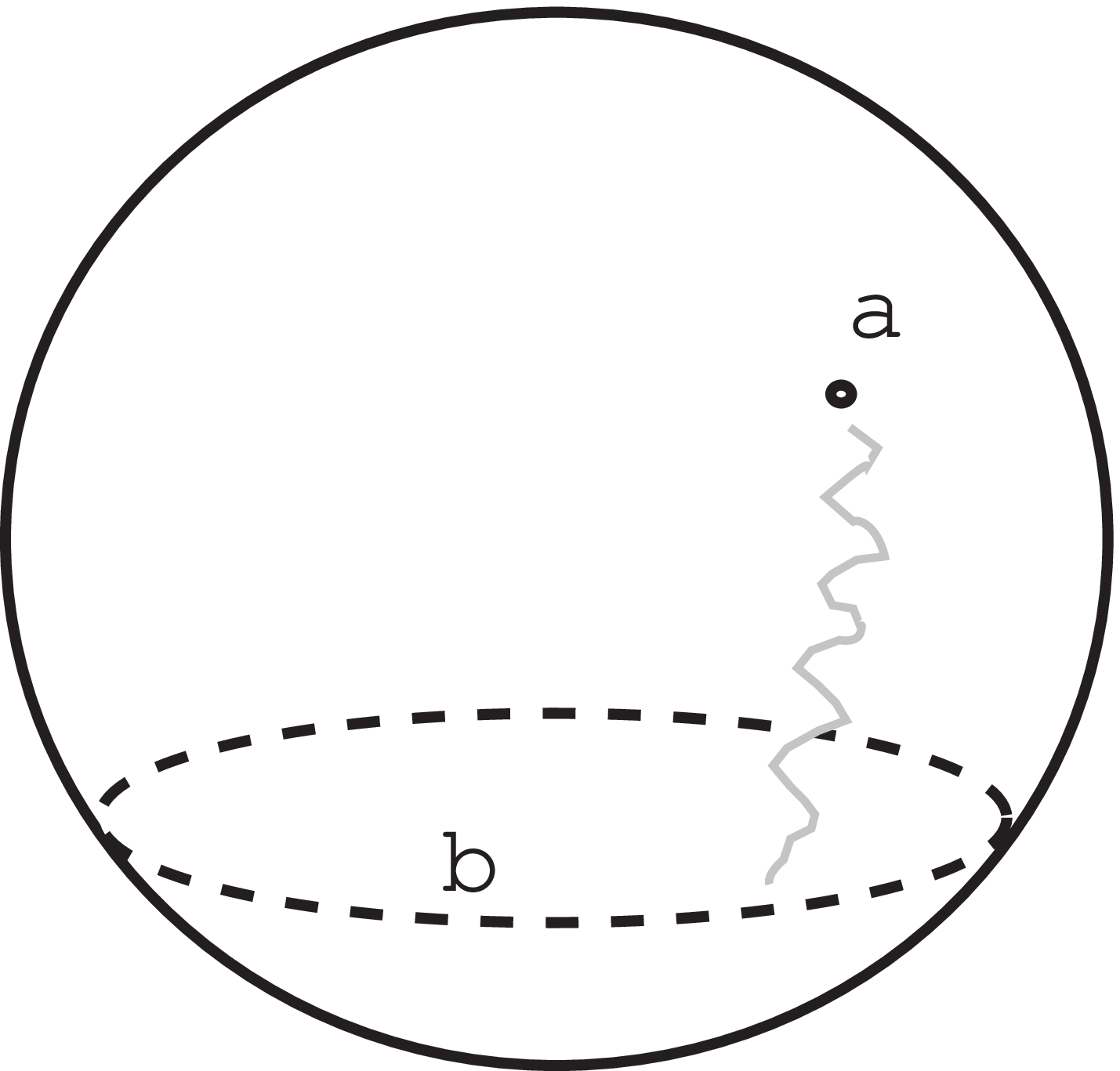}}
        \nobreak\bigskip
    {\raggedright\it \vbox{
{\bf Figure \Ddistance.}
{\it   Using a D0 brane as boundary condition $a$ we can probe for the 
location of brane $b$ by studying the lowest mass of the stretched strings.    } }}}}
    \bigskip}

Now, let us consider  the open string statespace $\CH_{ba}^{\rm open}$ with 
  $a$ corresponding to a rotated D0 brane, $a= g_L\cdot \vert B(\lambda=0)\rara$ and 
$b=\vert B(\lambda)\rara$. A typical string in this space may be pictured as 
in   Fig. \Ddistance. This picture suggests a way to  
 ``measure the distance'' between the two branes, and thereby to 
define the positions of branes in the spirit of 
\refs{\DouglasGE,\DouglasBE,\DouglasZW}.
The  
picture suggests that the open-string channel  partition function  
has an expansion for small $q_o$ 
\eqn\openpart{
{\Tr}_{\CH_{b,a}^{\rm open} } q_o^{L_0} 
~~ {\buildrel ? \over = } ~~
q_o^{(T_f D)^2 } + \cdots 
}
where $D$ is the   geodesic distance between the 
center of the   $D0$ brane   at $g_L$
and the brane $b$.  $T_f$ is the fundamental string tension (we set $\alpha'=1$ here, so 
$T_f = 1/(2\pi)$). 

We can actually compute the $q_o$ expansion of \openpart\ 
using  the expression for the boundary state together 
with  the Cardy condition: 
\eqn\rotcard{
{\Tr}_{\CH_{b,a} } q_o^{L_0 - c/24 } = 
\langle\langle B(\lambda)\vert q_c^{\half ( L_0 + \tilde L_0 - c/12)} \rho_L(g ) \vert B(0)\rara 
}
where 
$q_c = e^{2\pi i \tau_c}$, $q_o = e^{-2\pi i /\tau_c}= e^{2\pi i \tau_o}$. 
The computation is straightforward. Let us quote the result for $SU(2)$. 
If $g_L$ is conjugate to 
\eqn\ggee{
\pmatrix{ e^{i \chi}& 0 \cr 0 & e^{- i \chi}\cr} 
}
with $0 \leq \chi \leq \pi$, 
then, in the Ramond sector, the leading power of $q_o$ in \openpart\ is 
\eqn\leadingferm{
k \left( {\hat \chi_j - \chi \over 2\pi } \right)^2 .
}
Here $\hat \chi_j = \pi (2j+1)/k$, and the brane $b$ is 
labelled by $j\in \{0, \half, \dots, {k-2\over 2} \} $.  
The formula \leadingferm\ is {\it precisely} $(T_f D)^2$, 
as naively expected. 
Again, we see that the geometrical picture of the 
branes  is beautifully reproduced from 
the conformal field theory. 
\foot{In the bosonic case \leadingferm\ turns out 
to be $
{k+2\over 4} \left( {\hat \chi_j - \chi \over \pi } \right)^2 +\half 
{\chi\over \pi}(1-{\chi\over \pi} )- {1\over4( k+2)}$ 
so it is only for $k(\delta \chi)^2 \gg 1$ that the conclusion holds.} 
Very similar remarks hold for the D-branes in coset models 
\MaldacenaKY. 
 
\subsec{ Why are the branes stable? }

The geometrical picture advocated in the previous  sections raises an 
interesting puzzle. We will now describe this puzzle, and its 
beautiful resolution   in  \BachasIK\PawelczykAH.

Consider a D-brane wrapping    $\CO_{\lambda,k}\subset G$ once, as in 
Fig. \symmbrane. 
In the context of the type II string theory, the
brane has a nonzero tension $T$, with units of  energy/volume. Hence,  
wrapping a submanifold $\CW$ with a brane costs energy $E\sim T\vol(\CW)$. 
However, the  
regular conjugacy classes $\CO_{\lambda,k} \subset G$
are homologically trivial. For example, for $SU(2)$,  $\CO=S^2\subset S^3$. 
We therefore expect the brane to be unstable and to  contract to a point.

This leads to a paradox: 
We know from conformal field theory that the brane is 
absolutely stable. From the expression for the boundary 
state we can compute the spectrum of operators in the open 
string statestate from 
\eqn\a{
{\Tr}_{
\CH_{\lambda, \lambda} } q^{L_0 - c/24}
}
and we   find all $\Delta_i\geq 1$. According to 
\betafun\ it follows that 
 there are no unstable 
flows under $\beta$ away from this point! 

 The resolution of the paradox lies in the fact that   
 D-branes also have gauge theory degrees of 
freedom on them. The brane carries a  $U(1)$ line 
bundle $\CL \to \CO$ with connection.
 If this bundle is twisted then there 
is a stabilizing force opposing the tension.

To illustrate the resolution in the simplest terms, consider the example of $SU(2)$ with 
conjugacy class $\CO = S^2$, of radius $R=\sqrt{k}\sin \chi$.   
If the Chan-Paton line bundle of the brane has Chern class 
$n\in \IZ$, then 
$\int_{\CO} F = 2\pi n$. It follows that the Yang-Mills action is 
\eqn\ymi{
\int_{S^2} F\wedge * F \sim {n^2\over R^2} 
}
and hence we can 
evaluate the $g$-function \hrdbi\ 
\eqn\ymii{g(\chi) \sim R^2 + {n^2\over R^2} }
This has a minimum at   $\chi \sim \pi n/k $, and hence we 
expect an RG flow in the sector of $\CB$ determined by $n$ 
to evolve to this configuration. 
 
%
 
The above arguments have been generalized from $SU(2)$ to higher rank 
groups in  \MaldacenaXJ\BordaloEC. The result that emerges is that  
 $\vert B(\lambda) \rara$ can be pictured, semiclassically, as 
 wrapping the conjugacy class $\CO_{  \lambda,k}$. The brane 
is singly wrapped, and its  Chan-Paton line bundle $\CL_{  \lambda} \to \CO_{  \lambda,k}$ 
has first Chern class
\eqn\cpline{
c_1(\CL_{  \lambda}) =   \lambda + \rho \in H^2(G/T;Z) \cong \Lambda_{\rm weight}.
}
(For further details see \MaldacenaXJ.) 

It is interesting to study the $g$-function and its approximation by 
the DBI action in this problem. We restrict attention to the bosonic 
WZW model. 
Let $\chi$ parametrize the conjugacy classes in $G$. For the 
Chan-Paton line bundle \cpline\  the DBI action 
\eqn\dbiwzw{
g_{DBI}(  \chi) := \int_{\CO_\chi} \sqrt{\det(g+ F+B)} 
}
as a function of $\chi$ is minimized at   $ \chi_* = 2\pi (\lambda + \rho)/(k+h)$, 
where it takes the value: 
\eqn\wbwed{
g_{DBI}(\chi_*)/g_{DBI}(0) = \prod_{\alpha>0 } \biggl({k \sin \half \alpha \cdot \chi\over \pi 
\alpha\cdot \rho}\biggr).
}
Here the product is over positive roots. 
This compares remarkably well with the   exact CFT answer:
\eqn\axct{
g(\lambda)/g(0) =  
\prod_{\alpha>0 } \biggl({\sin \pi \alpha \cdot (\lambda+\rho)/(k+h) \over 
\sin \pi \alpha \cdot  \rho/(k+h)  }\biggr).
}
Note that the right-hand side   is the quantum dimension $d_q(\lambda)$, in harmony with 
\bdryent\  above.

\subsec{How collections of D0 branes evolve to symmetry-preserving branes}

The semiclassical picture of the symmetry-preserving 
branes we have just described 
raises  an important new point. In type II 
string compactification, if a brane carries a topologically 
nontrivial Chan-Paton bundle then it carries nontrivial 
induced D-brane charge. In the present case since the 
Chan-Paton line bundle has   $\int_{\CW} e^{c_1(\CL)}\not=0$  
it carries D0 charge. This suggests that the conformal 
fixed point characterized by $\vert B(\lambda) \rara$ is 
in the same component of $\CB$ as the fixed point corresponding 
to a   ``collection of D0 branes.''  In this section we 
review why that is true. 
\foot{Actually, the most obvious embedding of the $SU(2)$ WZW model into 
a type IIA background using a Feigin-Fuks superfield 
produces a background for which 
the definition of RR D0 charge is in fact subtle. The 
relevant $U(1)$ RR gauge group is spontaneously broken to $\IZ_k$ 
due to the condensation of a spacetime scalar field of charge $k$.  
See \MaldacenaSS\ for more discussion. }

By a ``stack of D0 branes'' physicists mean the boundary state 
$L \vert B(0)\rara$ for some positive integer $L$. 
By definition, the open string sectors for such a stack of 
D0 branes have state spaces 
\eqn\stackdo{
\CH_{L B(0) , b }^{\rm open} = \IC^L \otimes \CH_{B(0),b}^{\rm open}
}
for any boundary condition $b$. 

{\it Claim}:  If   $\lambda\in P_k^+$ and $L=d(\lambda)$, then
 $L \vert B(0)\rara$ is in the same component of 
$\CB$ as $\vert B(\lambda)\rara$.

Note that  
\eqn\bdryratio{
{g(B(\lambda))\over g(LB(0) )} = {S_{\lambda,0}\over L S_{00}} = {d_q(\lambda)\over d(\lambda)} < 1
}
so the claim is nicely consistent with the $g$-theorem. In particular, if 
these fixed points can be connected by RG flow then  
 $L$ D0's are unstable to $\lambda$, and not vice versa. We sometimes 
refer to this instability as the ``blowing up effect.'' 

The RG flow in question arises in the   theory of the Kondo effect
and was studied by Affleck and Ludwig \AffleckTK\AffleckNG.  
Their results  were applied  in the present 
context by Schomerus and collaborators.
 See \SchomerusDC, and references therein. 
Kondo model trajectories are obtained by 
perturbing a conformal fixed point by the 
holonomy of the unbroken current algebra in some 
representation. The flow, which should take 
 $L \vert B(0)\rara$ to $\vert B(\lambda)\rara$, 
is given by considering the disk partition function 
\eqn\bdrybos{
Z(u) = \langle {\Tr}_{\lambda}\biggl( P \exp \oint d\tau u J(\tau) \biggr)\rangle.
}
As  explained in \SchomerusDC, the results of Affleck and Ludwig 
lend credence to the main claim.

Actually, it is important to take into account $\CN=1$ supersymmetry in 
this problem. 
\foot{The following argument combines elements from 
\refs{\SchomerusDC,\HikidaPY,\MaldacenaXJ}. } In the supersymmetric 
WZW model we have a superfield 
\eqn\spfld{
{\bf J}_a(z) = \psi_a(z) + \theta I_a(z),
}
where we have chosen an orthonormal basis for the Lie algebra, 
labelled by $a=1, \dots, \dim G$. The OPE's are \refs{\fqs,\knizhnik,\kac}
\eqn\superwzw{ \eqalign{ I_a(z) I_b(w) & \sim {k
\delta_{ab}\over (z-w)^2} +  f_{ab}^{~~c} {I_c(w)\over z-w}
+\cdots \cr I_a(z) \psi_b(w) & \sim {f_{ab}^{~~c}
\psi_c(w)\over z-w} + \cr \psi_a(z) \psi_b(w) & \sim {k
\delta_{ab}\over (z-w)} +\cdots\cr} }
By a standard argument the currents $J_a = I_a + {1\over 2k} f_{abc} \psi_b \psi_c$
decouple from the fermions and satisfy a current algebra with 
level $k-h$. The Hamiltonian and supersymmetry charge are given 
(in the Ramond sector) by 
\eqn\susychge{
\eqalign{
Q & = \oint dz  \biggl( {1\over k} J_a \psi_a - {1\over 6k^2} f_{abc} \psi_a \psi_b \psi_c \biggr) \cr
H & ={1\over 2k} \oint dz  \left( :J_a J_a: +  \p \psi_a \psi_a \right) \cr}
}
The supersymmetry transformations are $[Q,\psi_a] = I_a $ and 
$[Q, I_a] = \p \psi_a $. 

Now, let us add a Kondo-like boundary perturbation preserving $\CN=1$ supersymmetry. 
This is given by choosing a representation $\lambda$ of $G$ and taking 
\eqn\bdryferm{
g(u) = \langle {\Tr}_{\lambda}\biggl( P \exp \oint d\tau u I(\tau) \biggr)\rangle
}
Using $[Q, I_a] = \p \psi_a $ to vary the perturbed action in \bdryferm\ 
we may compute the perturbed supercharge $Q_u$. This operator acts 
on  $\IC^L \otimes \CH^{\rm open}$ as
\eqn\newspchge{
\eqalign{
Q_u & = Q+ u \psi^a(0)  S^a\cr
&  = Q+ u \sum_{n\in \IZ}  \psi^a_n S^a \cr}
}
In the first line  we have passed to a Hamiltonian formalism for the 
open string  on a space 
$[0,\pi]$ (and we are only modifying the boundary condition at $\sigma=0$), 
and we have introduced explicit generators $S^a$ for the finite dimensional 
representation $\IC^L$ of the group $G$. In the second line we have used the 
doubling trick to express the result in terms of modes of a single-valued 
chiral vertex operator on the plane $\IC$. 
We can now compute the perturbed Hamiltonian 
\eqn\prtbd{
H_u = Q_u^2 = H + u I^a(0)S^a + u^2 (\psi^a(0) S^a)^2 
}
The third term in \prtbd\ is singular, but the renormalization of this 
term is fixed by the requirement of supersymmetry. The Hamiltonian 
can be written  as 
\eqn\newhamii{
\eqalign{
& H_u= {1\over 2k} \sum_n \left( : (J^a_n + uk S^a)(J^a_{-n} + uk S^a) : + n \psi^a_n \psi^a_{-n} \right)  \cr
& + \half u (u-{1\over k}) \sum_{n,m} f^{abc} \psi_n^a \psi_m^b S^c \cr}
}
so that the vacuum of the theory evolves in a complicated way
as a function of $u$. Note that, 
exactly for $u=u_*=1/k$,  the Hamiltonian 
simplifies into 
\eqn\newham{
H_* = {1\over 2k} \sum_n \left( : \CJ^a_n \CJ^a_{-n} : + n \psi^a_n \psi^a_{-n} \right) 
}
where 
\eqn\newcrrt{\CJ^a_n  := J^a_n + S^a} 
also satisfy a current algebra with level $k-h$. Thus, at $u=u_*$, we can build a 
new superconformal algebra with these currents.

The previous paragraph strongly 
suggests that $u_*=1/k$ is a second critical point for the 
boundary conformal field theory. Now, we can use an 
observation of Affleck and Ludwig. If $\CH_{\lambda'}$ is 
a representation of $J_n^a$, then  
with respect to a new current algebra $\CJ_n^a$ we 
can decompose:  
\eqn\decomp{
\IC^L\otimes \CH_{\lambda'} \cong \oplus_{\lambda''\in P^+_k} N_{\lambda, \lambda'}^{\lambda''} \CH_{\lambda''}
}
(An easy way to prove \decomp\ is to   consider the cabling of   Wilson lines in 3D Chern-Simons theory, 
and use the Verlinde algebra.) 
Therefore it follows that  
\eqn\bdryidnt{
{\Tr}_{\lambda_1}\biggl( P \exp \oint   u_*  I   \biggr) \vert B(\lambda_2)\rara = 
\sum_{\lambda_3} N_{\lambda_1,\lambda_2}^{\lambda_3} \vert B(\lambda_3) \rara
} 
where the boundary states on the RHS are constructed using 
$\CJ_n^a$. It would be worthwhile to give a direct proof of 
\bdryidnt. The     identity has been verified at large $k$ 
in \HikidaPY.

Let us close this subsection with a number of remarks. 

\item{1.} Note that when there is more than one term on the 
right-hand side of \decomp\ a local boundary condition has 
evolved into a (mildly) nonlocal boundary condition. Regrettably, 
this muddies the proposed definition of D-branes as local boundary 
conditions preserving conformal invariance.

\item{2.} The instability of a stack of $L$ $D0$-branes to decay to a
symmetry-preserving brane has been much discussed in the literature
in  the framework 
of noncommutative gauge theory. See \SchomerusDC\ and references therein. 
The arguments show that the ``D-brane instantons'' of the previous 
section should be viewed as real-time processes taking place in 
classical string theory.

\item{ 3. } The Kondo flows are integrable flows. The $g$ function 
has been studied in   \refs{\BazhanovFT,\BazhanovDR,\BazhanovDQ,\andrei,
\FendleyKJ,\LesageQF}  
in several examples and for certain boundary conditions related to the 
free fermion construction of current algebras.  It is possible that the techniques of 
\refs{\BazhanovFT,\BazhanovDR,\BazhanovDQ}  can be used to give exact results for how 
the boundary state evolves along the RG trajectory. This could 
be very interesting indeed.  

\item{4.} The ``blowing up effect'' is closely related to some work of 
\MickelssonJX\freedtalk.  These authors study families of Fredholm 
operators over the space of gauge fields on $S^1$. 
The perturbed supersymmetry operator along the RG flow is 
related to the family of Fredholm operators studied in 
\MickelssonJX\freedtalk.

\item{5.} One of the most remarkable aspects of the blowing-up
effect is the disappearance of $k$ D0-branes ``into nothing.'' 
Let us stress that this is an effect   studied in 
the laboratory!
One studies electrons coupled to a magnetic ``impurity.'' Translating 
this system into conformal field theory terms \AffleckGE\andrei\  reveals 
the boundary $SU(2)$ model with $k=1$; the presence of 
the magnetic impurity, in the high temperature regime, 
 translates into the presence of a single 
D0 brane. The RG flow parameter is the temperature, and, as 
$T \to 0$, the magnetic impurity is screened and ``disappears.'' 
The absence of the magnetic impurity corresponds to the disappearance of 
the D0 brane.

\item{6.} The effect we are discussing can be related, by 
U-duality, to the Myers effect \MyersPS. (Apply $S$-duality 
to a IIB solitonic 5-brane.) 

\item{7.} Actual evaluation of the standard D0-brane charge 
formula $\int_{\CO_{\lambda,k}} e^{F+B} $ yields a quantum 
dimension for the group. As far as we know, 
this curious fact has not yet been 
properly understood. 

\subsec{ The $D0$ charge group  } 

At this point we have {\it two} notions of D0-brane charge. 
On the one hand, we have naive  D0 charge $L= d(\lambda)$.
On the other hand, as we explained in     
section 4, due to D-brane instantons, the true  D0 charge, which 
we denote by  $q(\lambda)$,
must be a torsion.   Indeed, we  
know that D-brane instanton effects will impose a relation:  
\eqn\twoch{
q(\lambda) = d(\lambda) {\rm mod}~ d_{k,N}
}
for some integer $d_{k,N}$, 
but, (without Hopkins), 
it is hard to account for all possible D-brane instantons. 
So, we will now determine 
the order, $d_{k,N}$, of the torsion group using the blowing up 
effect and a simple observation regarding rotated branes.

\bigskip
{\vbox{{\epsfxsize=1.1in
        \nobreak
    \centerline{\epsfbox{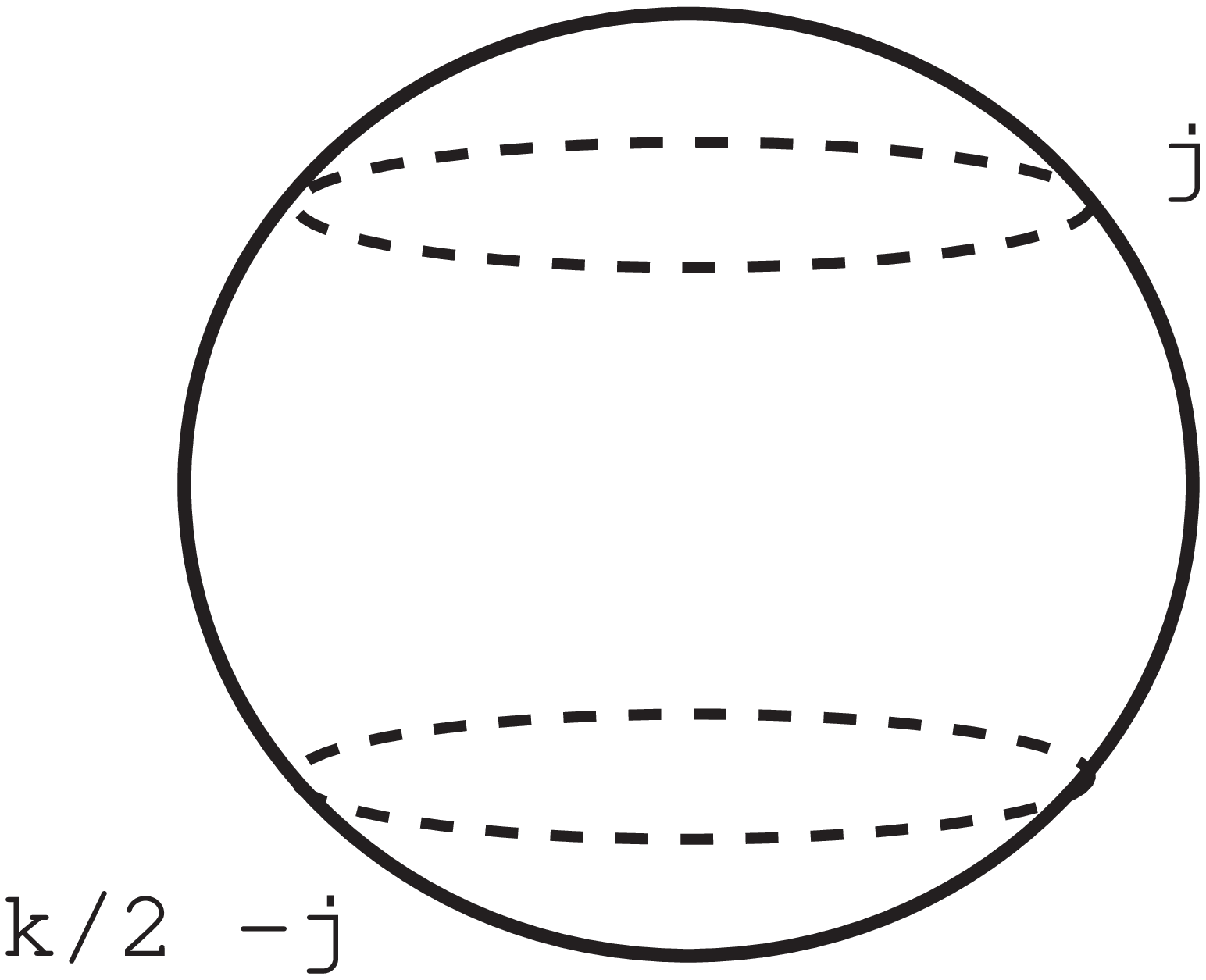}}
        \nobreak\bigskip
    {\raggedright\it \vbox{
{\bf Figure \rotated.}
{\it    Two symmetry-preserving branes related by a rigid rotation.     } }}}}
    \bigskip}

Recall from section 5.3 that we can rotate our branes by $G_L \times G_R$. 
Sometimes it can happen that the special 
conjugacy classes can be rotated into one another 
\eqn\rtsby{
g_L \CO_{\lambda,k} g_R = \CO_{\lambda',k} .
}
For example, if  $G=SU(2)$ the conjugacy classes 
$\CO_{j,k}$ and $\CO_{k/2 - j, k}$ can be rotated into each 
other: 
\eqn\a{
(-1)\cdot \CO_{j,k} = \CO_{\half k - j, k}
}
as in Fig. \rotated. Let us ask  
which representations are related in this way. 

In order to answer this question we use the  
 well-known relation between the center of a 
compact  connected, simply-connected Lie group $G$ and the 
automorphisms of the extended Dynkin diagram. 
For example, if $G=SU(N)$, $Z(G)\cong Z_N$, and $Z_N$ 
acts on the extended Dynkin diagram by rotation. 
Next,    the automorphisms of the 
extended Dynkin diagram act on the space of 
level $k$ integrable representations $P^+_k$. 
For example, for $SU(2)$ 
\eqn\spflw{  j \to j'= \half k-j  ,}
while for 
  $\widehat{SU(N)}_k$ the generator of $Z(G)$ acts on the 
Dynkin labels by 
\eqn\a{
 \lambda = (a_1,\dots, a_{N-1}) \to \lambda' = (k-\sum a_i , a_1, \dots, a_{N-2}) .
}
A beautiful result of group theory   is that if  $z\in Z(G)$ then   
\eqn\zeeconj{ 
z\CO_{\lambda,k} = \CO_{z\cdot \lambda ,k}
}

Now we can use \zeeconj\ to determine the order of the D0 charge 
group. To see this note first that 
    two  branes related by a rigid rotation must have the 
same D0 charge!
On the other hand, $ \lambda$ and $z\cdot \lambda$ 
have {\it different } dimensions, and hence have 
different naive D0 charge.
Therefore, we seek an integer $d_{k,N}$ such that 
\eqn\a{
d(z\cdot \lambda) = \pm d(\lambda) {\mod} ~ d_{k,N} \qquad \forall z\in Z(G), \lambda\in P^+_k
}
where  the sign  $\pm $ depends on $z$ and the rank of $G$, and  accounts for  orientation
(see  \MaldacenaXJ\  for more details).  It turns out that this  
condition determines $d_{k,N}$: 
\eqn\a{
d_{k,N}  = {\rm gcd}\bigl[{k\choose 1}, {k\choose 2},
\dots, {k \choose N-1}\bigr].
}
in perfect agreement with Hopkins' result! 
(See \FredenhagenEI\MaldacenaXJ\ for details of some of the 
arithmetic involved. )

Remarks: 

\item{1.} The generalization of $d_{k,N}$ to other compact
simple Lie groups has been discussed in \BouwknegtBQ.

\item{2.} 
After my talk at the conference, M. Hopkins made a 
curious remark which I  would like to record here. 
There is a simple mathematical relation between twisted 
{\it equivariant} K-theory and twisted K-theory of $G$ which 
has several of the same ingredients as the physical discussion 
we have just given.  
If $\pi_1(G)$ is torsion free   the Kunneth formula of 
\snaith\hodgkin\ suggests 
that the two twisted K-theories are related by  
\eqn\kunneth{
\IZ \otimes_{R(G)} K_{G,H}(G) = K_H(G)
}
Here the representation ring $R(G)$ is 
to be thought of as the ring of functions on 
the representation variety $G/T$ with $T$ acting by conjugation. 
$R(G)$ acts on $1\in \IZ$ by the dimension of the representation, 
while $K_{G,H}(G)$ is the Verlinde algebra, thanks to   the 
theorem of Freed, Hopkins, and Teleman \FreedJD\FreedMZ\fht.  
Curiously, from the point of view of algebraic geometry 
this means that the special conjugacy classes have an 
intersection with the identity element, when considered as 
varieties over $\IZ$.

\subsec{Comment on Cosets} 

The point of view explained above has potentially interesting 
applications to branes in coset models. Roughly speaking, if $L\subset G$ 
is a subgroup then the branes in the $\CN=1$ supersymmetric coset model 
$G/L$ should be classified by the twisted equivariant K-theory 
$K_{L,H}(G)$, where the twisting comes from the WZW $G$-theory. 
The branes in such coset models have been studied in many papers. 
See  
\refs{\LercheIV , \MaldacenaKY,\GawedzkiYE,\FalcetoEH
\FredenhagenKW,\FredenhagenQN,
\FredenhagenXF} for a sampling. 
For the $SU(2)/U(1)$ model the stable $A$-banes described in 
 \MaldacenaKY\  are in perfect accord with the twisted equivariant $K$-theory. 
The higher rank situation is somewhat more subtle and is 
currently under study by S. Schafer-Nameki \schafernameki.

\newsec{Conclusion}

Our goal in this talk was not to establish 
rigorous mathematical theorems but to explain how 
physics can suggest some intuitions for K-theory 
which are complementary to the more traditional  (and rigorous!) 
approaches to the subject.  Such alternative 
viewpoints and heuristics can sometimes suggest new 
and surprising directions for enquiry, or can suggest 
simple heuristics for already-known results. 
The above ``derivation'' of   the twisted K-theory 
of $SU(N)$ is just one example, 
but there are others. For example, the symmetry-preserving 
branes are precisely the branes which 
descend to branes in the $G/G$ gauged WZW model. 
The reason is that the gauge group acts on $G$ by 
conjugation, and only the symmetry-preserving boundary 
conditions preserve this gauge symmetry. Now, the 
$G/G$ WZW model is a topological field theory whose 
Frobenius algebra is the Verlinde algebra. This provides 
a simple perspective on the physics underlying the 
result of Freed, Hopkins, and Teleman \FreedJD\FreedMZ\fht.
(This remark is also related to the discussion of \FreedJD.)

Let us conclude by mentioning     
some future directions which might prove to be interesting
to the mathematics community, and which are suggested by 
the more physical approach to K-theory advocated 
in this paper.

First, in the context of spacetime supersymmetric models 
a special class of boundary conditions, the so-called 
``BPS states'' might have an interesting {\it product} 
structure \HarveyGC\MooreAR. Thus, perhaps the category 
of boundary conditions (or an appropriate subcategory) 
can also be given the structure of a tensor category. 

Second, the   RG approach to D-branes suggests an 
interesting generalization of the McKay 
correspondence to {\it non-crepant} toric resolutions 
of orbifold singularities \MartinecWG.  

Finally, the K-theoretic classification of D-branes in 
type II string theory must somehow be compatible 
with the U-duality symmetries these theories enjoy, 
and must somehow be compatible with   11-dimensional 
M-theory.  Only bits and pieces of this story are 
at present understood. It is   possible that 
the full resolution will be deep and 
will have interesting mathematical applications.


\bigskip  
\noindent{\bf Acknowledgements:}   
I would like to thank my collaborators on the work 
which was reviewed above, E. Diaconescu, D. Kutasov,  J. Maldacena,
M. Mari\~no, N. Saulina, G. Segal, N. Seiberg, 
and E. Witten. I have learned much from them. I am also indebted to 
N. Andrei, I. Brunner,  M. Douglas, D. Freed, D. Friedan, E. Getzler, 
M. Hopkins, S. Lukyanov, H. Saleur, S. Schafer-Nameki, 
V. Schomerus, S. Shatashvili, 
C. Teleman, and A. Zamolodchikov for useful discussions 
and correspondence
on this material. I would like to thank D. Freed for many 
useful comments on the draft. Finally,  I would also like to thank the Isaac Newton Institute 
for hospitality while this talk was written, and U. Tillmann for the 
invitation to speak at the conference. This work  
 is supported in part by DOE grant DE-FG02-96ER40949.

\listrefs

\bye